\documentclass[12pt]{article}
\usepackage{color}
\usepackage{amssymb}
\usepackage{amsmath}
\usepackage[dvips]{graphicx}
\usepackage{mathrsfs}
\usepackage[mathscr]{eucal}
\usepackage{caption}
\usepackage{subcaption}
\usepackage[dvipdfm]{hyperref}
\newcommand{\bs}{\boldsymbol}
\author{Mariusz Adamski\textsuperscript{1,}\footnote{Corresponding author: M. Adamski, e-mail:
mariusz.adamski@ift.uni.wroc.pl}, Janusz J{\c{e}}drzejewski\textsuperscript{1}
and Taras Krokhmalskii\textsuperscript{2}\\
\textsuperscript{1}Institute of Theoretical Physics, University of Wroc\l aw,\\
pl. Maksa Borna 9, 50--204 Wroc\l aw, Poland\\
\textsuperscript{2}Institute for Condensed Matter Physics,\\
1 Svientsitski Street, 79011 Lviv, Ukraine}

\title{Quantum-critical scaling of fidelity in 2D pairing models}

\begin{document}
\maketitle

\begin{abstract}
\noindent
The laws of quantum-critical scaling theory of quantum fidelity, dependent on the underlying system dimensionality $D$, have so far been verified in exactly solvable $1D$ models, belonging to or equivalent to interacting, quadratic (quasifree), spinless or spinfull, lattice-fermion models.
%\cite{rams PRL 11},\cite{rams PRA 11},\cite{ajk-2}
The obtained results are so appealing that in quest for correlation lengths and associated universal critical indices $\nu$, which characterize the divergence of correlation lengths on approaching critical points, one might be inclined to substitute the hard task of determining an asymptotic behavior of a two-point correlation function by an easier one, of determining the quantum-critical scaling  of the quantum fidelity.
However, the role of system's dimensionality has been left as an open problem. Our aim in this  paper is to fill up this gap, at least partially, by verifying the laws of quantum-critical scaling theory of quantum fidelity in a $2D$ case. To this end, we study correlation functions and quantum fidelity of $2D$ exactly solvable models, which are interacting, quasifree, spinfull, lattice-fermion models. The considered $2D$ models exhibit new, as compared with $1D$ ones, features:
at a given quantum-critical point there exists a multitude of correlation lengths and multiple universal critical indices $\nu$, since these quantities depend on spatial directions, moreover, the indices $\nu$ may assume larger values. These facts follow from the obtained by us analytical asymptotic formulae for two-point correlation functions.
In such new circumstances we discuss the behavior of quantum fidelity from the perspective of quantum-critical scaling theory. In particular, we are interested in finding out to what extent the quantum fidelity approach may be an alternative to the correlation-function approach in studies of quantum-critical points beyond 1D.
\end{abstract}

\section{\label{intro} Introduction}

In recent years, quantum phase transitions and quantum-critical phenomena constitute a subject of great interest and vigorous studies in condensed matter physics. Both, experimental and theoretical developments point  out to the crucial role that quantum phase transitions play in physics of frequently studied high-$T_c$ superconductors, rare-earth magnetic systems, heavy-fermion systems or two-dimensional electrons liquids exhibiting fractional quantum Hall effect \cite{sachdev QPTs}, \cite{vojta QPTs}. Quantum-critical phenomena have been also observed  in exotic systems as magnetic quasicrystals \cite{deguchi} and in artificial systems of ultracold atoms in optical lattices \cite{zhang}. The so called classical, thermal phase transitions originate from thermal fluctuations, a competition of internal energy and entropy, and are mathematically manifested as singularities in temperature and other thermodynamic parameters of various thermodynamic functions, and such characteristics of correlation functions as the correlation length, at nonzero temperatures. In contrast, quantum phase transitions originate from purely quantum fluctuations and are mathematically manifested as singularities in system parameters of the ground-state energy density, which is also  the zero-temperature limit of the internal energy density. Naturally, singularities of thermodynamic functions appear only in the thermodynamic limit. The importance of quantum phase transitions for physics and the related wide interest in such  transitions stems from the fact that, while a quantum phase transition is exhibited by ground states, hence often termed a zero-temperature phenomenon, its existence in a system exerts a great impact on the behavior of that system also at nonzero temperatures. A quantum-critical point gives rise to the so called quantum-critical region, which extends at nonzero temperatures,  in some cases up to unexpectedly high temperatures \cite{kopp chakravarty}, \cite{kinross}.

Theoretically, quantum phase transitions can be studied in quite complex quantum systems by qualitative and approximate methods, or in relatively simple but exactly solvable models by means of analytic methods and high-accuracy numerical calculations \cite{sachdev QPTs}. Naturally, for the purpose of testing and illustrating general or new ideas the second route is most suitable. Traditionally, this route involves studying the eigenvalue problem of a Hamiltonian, the ground state and excitation gaps, determining quantum-critical points and symmetries, constructing local-order parameters, calculating two-point correlation functions and their asymptotic behavior at large distances and in vicinities of quantum-critical points, with correlation lengths and the universal critical indices $\nu$ that characterize the divergence of correlation lengths on approaching critical points. Carrying out such a programme is a hard task, which has been accomplished only in a few one-dimensional models.
Among those models, there are quantum spin chains as the isotropic and anisotropic XY models in an external transverse magnetic field, including their extremely anisotropic version--the Ising model \cite{sachdev QPTs}. Only in one dimension those models are equivalent to lattice gases of spinless fermions, which can exactly be diagonalized, and exact results concerning the phase diagram, quantum-critical points, correlation functions and dynamics have been obtained (concerning XY model see \cite{lieb}, \cite{barouch I}, \cite{barouch II},  concerning the Ising model see \cite{pfeuty}, and for both models \cite{parkinson}). Needless to say that parallel results for a higher-dimensional model are desirable; this is the first motivation of our investigations presented in this paper.

In the last decade, fresh ideas coming from quantum-information science entered the field of quantum phase transitions. One of them is the so called quantum-fidelity method.
The carried out so far studies of particular models show that using this method it is possible to locate critical points
\cite{zanardi paunkovic 06}, \cite{barjaktarevic 08}, \cite{zhou 08} and to determine the correlation lengths and associated universal critical indices $\nu$ \cite{rams PRL 11}, \cite{rams PRA 11}, \cite{ajk-2}. This is achieved by studying (typically by numerical methods) quantum-critical scaling properties, with respect to the size of the system and the parameters of the underlying Hamiltonian, of the quantum fidelity of two ground states in a vicinity of a quantum-critical point. The index $\nu$ is extracted from numerically obtained plots of fidelity via the quantum-critical scaling laws of quantum fidelity, which have been derived by renormalization group arguments \cite{venuti zanardi 07}, \cite{schwandt 09}, \cite{albuquerque 10}, \cite{rams PRL 11}. The most comprehensive results concerning the fidelity approach have been obtained for one-dimensional quantum spin systems in a perpendicular magnetic field \cite{rams PRL 11} (the case of Ising model), \cite{rams PRA 11} (the case of XY model), \cite{ajk-2} (the case of a quasifree, pairing, lattice-fermion model). These results are very promising: except a vicinity of a multicritical point, the fidelity approach works fine. Since the task of determining quantum-critical scaling properties of the quantum fidelity is definitely much easier than the task of calculating large-distance asymptotic behavior of two-point correlation functions, one is tempted to consider the fidelity method as a substitute for the standard correlation-function approach. All that we have said above makes it desirable to verify the laws of quantum-critical scaling of fidelity and the effectiveness of fidelity approach in dimensions higher than one, where new features, not encountered in one-dimensional models may appear; this is the second motivation of our investigations reported in this paper. For that task we need an at least two-dimensional exactly solvable model, whose ground states, quantum-critical points with critical indices in their vicinities, correlation lengths, and analytic expressions for fidelity can be determined.

To go beyond the one-dimensional case, we consider lattice-fermion models which originate from the two-dimensional model of d-wave superconductivity proposed by Sachdev \cite{sachdev 02}(see also \cite{sachdev QPTs}), which are spinful pairing models represented by quadratic Hamiltonians. In many cases of physical interest the quadratic Hamiltonians are obtained by means of a mean-field approximation applied to quartic Hamiltonians of systems with two-body interactions and their parameters are related by self-consistency equations. While our Hamiltonians are also quadratic, their parameters are independent---not related by mean-field equations. General, mathematical considerations of some classes of such models, but without specifying hopping intensities or coupling constants, which therefore do not reach such subtleties as quantum-critical points or critical behavior of correlation functions, can be found in \cite{zanardi 07}, \cite{cozzini}. For translation-invariant hopping intensities and coupling constants the considered models are exactly solvable in any dimension, that is, in particular it is possible to derive analytical formulae for quantum fidelity and correlation functions of finite systems and then in the thermodynamic limit, where boundary conditions play no role. To limit further the great variety of possible models, we restrict the hopping intensities to nearest neighbors while the dimensionality is set to $D{=}2$.
The underlying lattice is chosen as a square one with hopping intensities invariant under rotations by $\pi/2$, while the interactions of our systems are required not to extend beyond nearest neighbors and to be either invariant under rotations by $\pi/2$ (the symmetric model) or to change sign after such a rotation (the antisymmetric model).

It is worth to mention here another class of models, which might be of interest in the context of this paper, known as reduced BCS models of superconductivity and superfluidity, see for instance \cite{dukelsky} and references quoted there. While those models are exactly integrable, analytical formulae for quantities of interest, such as correlation functions, are not available, except at the thermodynamic limit, where they coincide with the results of mean-field theories. For finite systems, all the quantities we are interested in, ground-states, quantum fidelity, ground-state correlation functions, are given in terms of only numerically available solutions of a set of coupled, nonlinear, algebraic equations, whose number amounts to the number of degrees of freedom. Therefore, despite some interesting features, for instance the ground state is not a BCS-like state, those models are not suitable for the kind of studies reported in this paper.

The general plan of the paper is as follows. In section \ref{models} we  define the two models studied in this paper, the symmetric model and the antisymmetric one, and give closed-form formulae for two basic ground-state two-point correlation functions, which are then used to define order parameters of those models. Later on, when specific asymptotic behavior of two-point functions is discussed, only the gauge invariant function, that is the offdiagonal matrix element of the ground-state one-body reduced density operator is taken into account. The purpose of the next section \ref{fidelity} is to present the quantum fidelity method of investigating quantum-critical points. In particular we provide the known quantum-critical scaling laws obeyed by fidelity in a vicinity of a critical point, and express the  fidelity of the models considered in the paper as a Riemann sum of an analytically given function of two variables -- the components of quasimomentum.
Then, in sections \ref{symm} and \ref{antisymm} we carry out our program of confronting predictions of quantum-critical scaling theory of fidelity with exact results obtained for 2D pairing models, the symmetric and antisymmetric ones. This program consists of two stages. Since a critical scaling theory is concerned with correlation lengths and critical indices, in the first stage analytic results for the spatial direction-dependent behavior of the  gauge-invariant two-point correlation function at sufficiently large spatial distances and sufficiently close to the quantum-critical points exhibited by our models are highly desirable. Having such results, one is able to directly infer exact expressions, in terms of system's parameters, for spatial direction-dependent correlation lengths and exact values of the corresponding critical indices. We provide such results in the paper; they are excerpts from our article \cite{ajk-1}, where comprehensive studies of the gauge-invariant two-point correlation function of our models have been carried out. In the second stage, we are concerned with the variations of quantum fidelity against the system's linear size or against the distance to a critical point, sufficiently close to a quantum-critical point. Having expressed the fidelity as a Riemann sum, we generate suitable high-accuracy numerical plots that reveal those variations. Then, our discussion concentrates on answering two questions. First, can we read off from those plots the values of the spatial direction-dependent correlation lengths? Second, can we identify regions where the behavior of fidelity matches the known critical scaling laws with the known exact values of related critical indices? Anticipating the results to be presented, we say only  that the answer depends on the kind of a critical point and the values of the associated critical indices $\nu$. As compared to the results of so far carried out analogous studies of $1D$ cases \cite{rams PRL 11},\cite{rams PRA 11},\cite{ajk-2}, the answer is more likely to be negative. Finally, in section \ref{summ}, we summarize our results and draw conclusions.

\section{\label{models} The models, their ground-states and ground-state correlation functions}

We consider a $D$-dimensional spinful fermion model, given by the Hamiltonian,
\begin{equation}
H {=}  \sum_{{\bs l},i,\sigma} \left[ \frac{t}{2}
\left(a^{\dagger}_{{\bs l},\sigma} a_{{\bs l}+{\bs e}_i,\sigma} + \mathrm{h.c.} \right)
- \frac{\mu}{D} a^{\dagger}_{{\bs l},\sigma} a_{{\bs l},\sigma}
- \frac{J}{2} \left(\sigma \Delta_i
%\left(
a^{\dagger}_{{\bs l},\sigma}a^{\dagger}_{{\bs l}+{\bs e}_i, -\sigma}
%a^{\dagger}_{{\bs l},-\sigma}a^{\dagger}_{{\bs l}+{\bs e}_i,\sigma} \right)
+ \mathrm{h.c.}\right)\right],
\label{ham1}
\end{equation}
where $a^{\dagger}_{{\bs l},\sigma}$, $a_{{\bs l},\sigma}$ stand for creation and annihilation operators, respectively, of a spin $1/2$ fermion, whose
spin projection on a quantization axis is $\sigma {=} \pm 1$ in units of $\hbar/2$, in a state localized at site ${\bs l}{=}(l_1,\ldots,l_D)$ of a $D$-dimensional hypercubic lattice. The edge of the lattice in the direction given by the unit vector ${\bs e}_i$, $i{=}1,\ldots,D$, whose $m$-th component is $\delta_{i,m}$, consists of $L_i$ equidistant sites, labeled by $l_i{=}0,1,\ldots,L_i{-}1$, $l_j{=}0$ for $j{\neq}i$. In all the considerations that refer to finite systems, special boundary conditions, specified below, are chosen.
The sums over ${\bs l},i$  in (\ref{ham1}) amount to  the sum over pairs of nearest neighbors, with each pair counted once. The real and positive parameter $t$ is the nearest-neighbor hopping intensity, $\mu$--the chemical potential, $J$--the coupling constant  of the gauge-symmetry breaking interaction, and $\Delta_i$, $i{=}1,\ldots,D$, stand for direction-dependent, in general complex, dimensionless parameters.
In distinction to \cite{sachdev 02}, \cite{sachdev QPTs}, where Hamiltonian (\ref{ham1}) together with equations relating  $\Delta_i$ with $\mu$ and $J$ was studied in a context of a mean-field approximation to a kind of $t$-$J$ model, here $\Delta_i$ are free parameters independent of $\mu$ and $J$. Naturally, we can express the parameters $\mu$ and $J$ in units of $t$, while the lengths of the underlying lattice in units of the lattice constant, preserving the original notation.

We note that Hamiltonian (\ref{ham1}) is not gauge invariant unless $J{=}0$. It is also not hole-particle invariant unless $\mu{=}0$ and $\Delta_i$, $i{=}1,\ldots,D$, are real. The latter condition can be assumed to hold without any loss of generality, since
Hamiltonian (\ref{ham1}) with any complex $\Delta$ is unitarily equivalent to that with $\Delta$ replaced by $|\Delta|$.

Imposing, independently in each direction ${\bs e}_i$, $i{=}1,\ldots,D$, periodic or antiperiodic boundary conditions,
Hamiltonian (\ref{ham1}) can be simplified by passing from the site-localized to the plane-wave basis labeled by suitable wave vectors (quasimomenta) ${\bs k}$ whose components are denoted $k_{i}$, $i{=}1,\ldots,k_D$,
\begin{equation}
H = \sum_{{\bs k},\sigma}\varepsilon_{\bs k} c^{\dagger}_{{\bs k},\sigma} c_{{\bs k},\sigma}
- J \sum_{{\bs k},i}   \cos{k_i} \left( \Delta_i   c^{\dagger}_{{\bs k},\uparrow}c^{\dagger}_{-{\bs k},\downarrow} + \mathrm{h.c.} \right),
\label{ham2}
\end{equation}
where $\varepsilon_{\bs k}$ stands for the dispersion relation of the hopping term,
\begin{equation}
\varepsilon_{\bs k} = \sum_i \cos k_i - \mu.
\label{epsilon}
\end{equation}
In what follows we set $L_i{\equiv}L$, with even $L$. Then, the components of wave vectors $\bs k$ can be chosen to assume the values
$k_i{=}\pi(2l_i{-}L_i{+}2)/L_i$ in the case of periodic boundary conditions and the values
$k_i{=}\pi(2l_i{-}L_i{+}1)/L_i$ in the case of antiperiodic boundary conditions.
Such Hamiltonians can readily be diagonalized by means of the Bogoliubov transformation. The dispersion relation of quasi-particles reads
\begin{equation}
E_{\bs k} + \sum_{\bs k} \left( \varepsilon_{\bs k} - E_{\bs k} \right),
\label{spectrum}
\end{equation}
where $\sum_{\bs k} \left( \varepsilon_{\bs k} - E_{\bs k} \right)$ is the ground-state energy, and $E_{\bs k}$, given by
\begin{equation}
E_{\bs k} =\sqrt{ \varepsilon_{\bs k}^2 + \left|J \sum_i \Delta_i \cos k_i \right|^2},
\label{Ek}
\end{equation}
are the single quasi-particle energies.
For a suitable choice of boundary conditions specified above, as long as our system is finite the excitation energies $E_{\bs k}$ remain strictly positive: $E_{\bs k} > 0$ for all values of ${\bs k}$, and this is assumed to hold in the sequel. Specifically, to avoid closing down
of the gap in finite systems, represented by the two-dimensional models discussed in the sections that follow, we impose periodic boundary conditions along one axis and antiperiodic ones along the orthogonal axis.

The Hamiltonian (\ref{ham1}) preserves parity; therefore without any loss of generality we can restrict the state-space to the subspace of even number of fermions. In this subspace, the state $|0\rangle_{qp}$ -- the quasi-particle vacuum  of an unspecified (but even) number of fermions, defined by
\begin{equation}
|0\rangle_{qp}= \prod_{\bs k}(u_{\bs k} + v_{\bs k} c^{\dagger}_{{\bs k},\uparrow}c^{\dagger}_{-{\bs k},\downarrow})|0\rangle,
\label{gs}
\end{equation}
where $|0\rangle$ is the fermion vacuum, with $u_{\bs k}$ real and positive,
\begin{equation}
u_{\bs k}= \sqrt{\frac{1}{2}\left(1+ \frac{\varepsilon_{\bs k}}{E_{\bs k}}\right)},
\label{uk}
\end{equation}
and, in general, complex $v_{\bs k}$,
\begin{equation}
|v_{\bs k}|= \sqrt{\frac{1}{2}\left(1- \frac{\varepsilon_{\bs k}}{E_{\bs k}}\right)}, \,\,\,
\arg v_{\bs k} = \arg \left( J \sum_i \Delta_i \cos k_i \right),
\label{vk}
\end{equation}
is the eigenstate of (\ref{ham2}) to the lowest eigenenergy, $\sum_{\bs k} \left(\varepsilon_{\bs k} - E_{\bs k}  \right)$.
As long as $E_{\bs k} > 0$ for all values of ${\bs k}$, the unique ground state $|0\rangle_{qp}$ is the vacuum of elementary excitations (quasi-particles).
However, on passing to the thermodynamic limit, when the system's linear sizes in all directions tend to infinity, the minimum of $E_{\bs k}$ over ${\bs k}$ (i.e.the excitation gap in the spectrum  of quasi-particles) may approach zero at special values of the chemical potential $\mu$ and the coupling constant $J$. Those special points in the $(\mu,J)$-plane, where $E_{\bs k}$, as a function of continuous wave vector ${\bs k}$, vanishes, constitute the quantum-critical points at which the system undergoes continuous quantum phase transitions.

All the correlation functions of considered systems can be expressed in terms of two basic two-point correlation functions. Since we are interested only in ground-state correlation functions, taking into account the lattice-translation invariance of our system
these two basic two-point correlation functions can be chosen as follows:
\begin{equation}
_{qp}\langle 0|a^{\dagger}_{{\bs 0},\sigma} a_{{\bs r},\sigma}  |0 \rangle_{qp}  \qquad \textrm{and} \qquad
_{qp}\langle 0|a_{{\bs 0},\sigma} a_{{\bs r},-\sigma}  |0 \rangle_{qp},
\label{corr G g}
\end{equation}
with some $\sigma$.
The first correlation function,
$ _{qp}\langle 0|a^{\dagger}_{{\bs 0},\sigma} a_{{\bs r},\sigma}  |0 \rangle_{qp}$, is gauge and spin-flip invariant;
for ${\bs r}\neq 0$ it represents offdiagonal matrix elements of the ground-state one-body reduced density operator.
Expressing $a^{\dagger}_{{\bs 0},\sigma}$, $a_{{\bs r},\sigma}$ by the creation and annihilation operators of quasi-particles
(the Bogoliubov transformation) we get
\begin{equation}
 _{qp}\langle 0|a^{\dagger}_{{\bs 0},\sigma} a_{{\bs r},\sigma}  |0 \rangle_{qp} =
 \frac{1}{L^D} \sum_{\bs k} |v_{\bs k}|^2 \exp i{\bs k}{\bs r} =
-\frac{1}{2L^D} \sum_{\bs k} \frac{\varepsilon_{\bs k}}{E_{\bs k}} \exp i{\bs k}{\bs r},
\label{corr_G 1}
\end{equation}
which, upon using the invariance of  $\varepsilon_{\bs k}$ and $E_{\bs k}$  with respect to reflections of ${\bs k}$ in coordinate axes, in the thermodynamic limit becomes
\begin{equation}
\lim_{L \to \infty} { _{qp}\langle } 0|a^{\dagger}_{{\bs 0},\sigma} a_{{\bs r},\sigma}  |0 \rangle_{qp}
\equiv G({\bs r}) =
- \frac{1}{2 \pi^D} \int_{[0,\pi]^{D}} d{\bs k} \frac{\varepsilon_{\bs k} }{E_{\bs k}}
\prod_{j=1}^D \cos k_j r_j \,.
\label{corr_G 2}
\end{equation}
Choosing the spin projection $\sigma= +1$, the second correlation function, measuring the degree of gauge-symmetry breaking, amounts to
\begin{equation}
_{qp}\langle 0|a_{{\bs 0},+} a_{{\bs r},-}  |0 \rangle_{qp} =
-\frac{1}{L^D} \sum_{\bs k} u^{*}_{\bs k} v_{\bs k} \exp -i{\bs k}{\bs r}=
-\frac{1}{2L^D} \sum_{\bs k} \frac{J\sum_i \Delta_i \cos k_i}{E_{\bs k}} \exp i{\bs k}{\bs r},
\label{corr h 1}
\end{equation}
which, by the above arguments, in the thermodynamic limit becomes
\begin{equation}
\lim_{L \to \infty} { _{qp}\langle } 0|a_{{\bs 0},+} a_{{\bs r},-}  |0 \rangle_{qp}
\equiv h({\bs r}) =
- \frac{1}{2 \pi^D} \int_{[0,\pi]^{D}} d{\bs k} \frac{J\sum_i \Delta_i \cos k_i}{E_{\bs k}}
\prod_{j=1}^D \cos k_j r_j.
\label{corr_h 2}
\end{equation}

Both the above defined two-point correlation functions are used to define the order parameters in the ground-state phase diagrams of the models considered below. The two-point correlation functions decay with increasing distance $|{\bs r}|$ between the points. In gapped phases their decay is dominated by an exponential factor, $\exp(-|{\bs r}|/\xi)$, which defines the correlation length $\xi$. If additionally $(\mu,J)$-points approach a quantum-critical point, i.e. the distance $\delta$ between them tends to zero, then $\xi$ diverges as $\delta^{-\nu}$, which in turn defines a universal critical index $\nu$ associated with a particular quantum-critical point. For $D{>}1$, two-point correlation functions, and hence correlation lengths $\xi$ as well as the critical indices $\nu$ may depend on spatial direction of vector ${\bs r}$ .

In what follows we shall study only the case of $D{=}2$. Clearly, all the above $D$-dimensional expressions can be adapted to the 2D case by setting $\Delta_i{=}k_i{=}r_i{=}0$ for $i>2$. Note however that even  with the restriction $D{=}2$, due to the freedom in choosing the relation between the parameters $\Delta_1$ and $\Delta_2$, formula (\ref{ham1}) represents a great variety of models. In this paper we limit our considerations to two cases only.  First we set $\Delta_1 {=} \Delta_2 {=} \Delta$, what results in the interaction term invariant under rotations by $\pi/2$; the corresponding model is dubbed symmetric. Then, we choose $\Delta_1 {=} - \Delta_2 {=} \Delta$, what results in the interaction term changing sign under rotations by $\pi/2$; the corresponding model is dubbed antisymmetric. In all the considerations below, that refer to the symmetric or the antisymmetric model, we make the identification $J|\Delta| {\equiv} J$. In both the cases the correlation functions of our systems are invariant not only with respect to lattice translations but also with respect to rotations by $\pi/2$.

In order to determine the correlation length $\xi$ and the critical index $\nu$ in a vicinity of some quantum-critical point we study the so called doubly-asymptotic behavior of the gauge-invariant correlation function $G({\bs r})$, defined in(\ref{corr_G 2}); the provided in this article analytical results are excerpts from  more comprehensive studies presented in \cite{ajk-1}. Clearly, for a fixed spatial direction and the parameters $\Delta_i$, $G({\bs r})$ depends on three parameters: $|{\bs r}|$ - the distance between the two points of the correlation function, the chemical potential $\mu$ and the coupling constant $J$. The so called doubly-asymptotic region is characterized as follows: for a fixed but sufficiently large $|{\bs r}|$ the $(\mu,J)$-points approach a quantum-critical point along a specific path.
The choice of those paths depends on the kind of considered critical points. The critical points of the symmetric model that are located at the $J$-axis, with $|J|{>}0$, and those of the antisymmetric model that are located on the lines parallel to the $J$-axis are approached along  $\mu$-paths that are parallel to the $\mu$-axis.
Then, the critical points of the symmetric model that are located at the $\mu$-axis are approached along  $J$-paths that are parallel to the $J$-axis. Finally, the multicritical point of the symmetric model is approached along the $45^{\circ}$-path.

As mentioned above, a novel feature of two-dimensional models, in comparison with the one-dimensional case, is that the two-point correlation function $G({\bs r})$ depends not only on the distance $|{\bs r}|$ but also on the direction of ${\bs r}$. Expressing ${\bs r}$ by its Cartesian coordinates, ${\bs r}{=}(r_1,r_2)$, we can parameterize directions by the ratio $r_1/r_2 {\equiv} n$.
Then, for a given critical point, we can expect $n$-dependent doubly-asymptotic behaviors of correlations. Unfortunately, the analytic asymptotic formulae for $G({\bs r})$ in offdiagonal directions (i.e. $n {\neq} 1$), which we have been able to obtain, apply only to points
${\bs r}$ such that $n {\geq} n_0 {>} 1$ or, by symmetry, $n {\leq} n^{-1}_0 {<} 1$, that is for offdiagonal directions which form a sufficiently small angle with the axial directions. Therefore, analytic asymptotic formulae in the diagonal direction (i.e. $n{=}1$) have been derived separately \cite{ajk-1}. These formulae define $n$-dependent correlation lengths $\xi^{(\pm)}_{\text{offdiag}}$ in offdiagonal directions satisfying the conditions specified above and the correlations length $\xi^{(\pm)}_{\text{diag}}$ in the diagonal direction, where the superscript plus refers to the symmetric model and minus -- to the antisymmetric one. For directions close to the diagonal one, that do not satisfy the above conditions, the correlation lengths and related critical indices have been determined numerically.
Interestingly, our analytical and numerical results show that, for each critical point of the symmetric  or the antisymmetric model, there are only two kinds of universal critical indices $\nu$: $\nu_{\text{offdiag}}$ for all offdiagonal directions and $\nu_{\text{diag}}$ for the diagonal direction.

\section{\label{fidelity} The ground-state fidelity and quantum-critical scaling laws}

Let ${\bs \lambda}$ be a vector whose components are those parameters of the considered system's Hamiltonian that drive a quantum phase transition, and ${\bs e}$ - a unit vector in the space of those parameters. Then, on varying parameter $\delta$ the vectors ${\bs \lambda} + \delta {\bs e}$ scan a neighborhood of ${\bs \lambda}$ along direction $\bs e$. For given ${\bs \lambda}$ and $\delta$, the ground-state fidelity at ${\bs \lambda}$ in direction $\bs e$,
${\mathscr{F}}_{\bs e}({\bs\lambda}, \delta)$, is the absolute value of the overlap of the ground states $|{\bs \lambda} \pm \delta {\bs e} \rangle$  at the points ${\bs \lambda} \pm \delta {\bs e}$,
\begin{equation}
{\mathscr{F}}_{\bs e}({\bs\lambda}, \delta) =
|\langle {\bs\lambda} - \delta {\bs e}|{\bs\lambda} + \delta {\bs e} \rangle|.
\label{qfidelity}
\end{equation}
A list of general, system independent, properties of ${\mathscr{F}}_{\bs e}({\bs\lambda}, \delta)$ can be found in \cite{gu 10}.
The transition point of a continuous quantum phase transition, i.e. a quantum-critical point,
denoted ${\bs \lambda}_c$, is characterized by the power-law divergence of the correlation length $\xi({\bs\lambda})$, as the quantum-critical point is approached:
$\xi({\bs\lambda}) \sim |{\bs \lambda} - {\bs \lambda}_c|^{-\nu}$, with $\nu$ being one of universal characteristics of a critical point. Alternatively, ${\bs \lambda}_c$ can be defined as the point where the gap between the ground-state energy and the energy of the lowest excited state vanishes. In reference \cite{zanardi paunkovic 06} it was demonstrated that the quantum-critical point ${\bs \lambda}_c$ can be identified as the value of ${\bs \lambda}$ at which the minimum of fidelity is realized. However, the fidelity approach seeks an answer to a more general question: does the behavior of quantum fidelity in a neighborhood of a quantum-critical point encode not only the location of that point but also some universal properties of the underlying quantum phase transition?
First results, pointing towards a positive answer to the raised question by providing some finite-size critical scaling of fidelity, have been obtained by Venuti and Zanardi \cite{venuti zanardi 07}.

Scaling theories of criticality are based on the assumption that the behaviour of a system is governed by a unique length---the correlation length. In particular, according to finite-size scaling theories, the properties of a system are close to those at the thermodynamic limit, if the linear size of the system, $L$, is much greater than the correlation length $\xi({\bs\lambda})$. In literature such a system (or regime) is called the thermodynamic limit system (regime) or off-critical system (regime). In the opposite case one speaks of a finite system (regime) or a quasicritical system (regime). However, discussing finite-size scaling properties of ground-state fidelity,
${\mathscr{F}}_{\bs e}({\bs\lambda}, \delta)$, we deal with two ground states of a system. Then, one of them may correspond to a quasicritical or critical regime while the other to an off-critical one. In the context of the quantum-critical-scaling theory of fidelity, it is expected that the role of the unique characteristic length of the system, that differentiates between a quasicritical system and an off-critical one, is played by the effective correlation length
${\tilde{\xi}}_{\bs e}({\bs\lambda}, \delta)$, which is defined as the smaller of the two correlation lengths
$\xi({\bs \lambda} \pm \delta {\bs e})$ \cite{rams PRA 11}. To turn attention to this fact and to emphasize that in all the considerations below the systems are finite, we introduce a somewhat nonstandard terminology. If $L{\gg}{\tilde{\xi}}_{\bs e}({\bs\lambda}, \delta)$, we call the system macroscopic; in the opposite case the system is called small. The crossover between small-system and  macroscopic-system regimes occurs, when the linear size of a system, $L$, satisfies the crossover condition:
\begin{equation}
L/{\tilde{\xi}}_{\bs e}({\bs\lambda}, \delta) \sim 1 .
\label{crossover}
\end{equation}

There are numerous papers devoted to critical scaling of small-system fidelity,  see \cite{gu 10}, \cite{albuquerque 10}, and references quoted there. Typically, small-system fidelity can be Taylor-expanded in $\delta$,
\begin{equation}
{\mathscr{F}}_{\bs e}({\bs\lambda}, \delta) =
1 - \frac{ \delta^2}{2} \chi_{\bs e}({\bs\lambda}) + \ldots ,
\label{susceptibility}
\end{equation}
where the first order term vanishes because of the symmetry of fidelity in $\delta$ at zero. The  coefficient of the second order term, $\chi_{\bs e}({\bs\lambda})$, is known as the fidelity susceptibility. One expects some universal scaling properties of $\chi_{\bs e}({\bs\lambda})$, provided $\bs\lambda$ is sufficiently close to a quantum-critical point ${\bs\lambda}_c$, where the correlation  length diverges:
$\xi({\bs \lambda}_c \pm \delta {\bs e})\sim |\delta|^{-\nu}$.
Fairly general, model-independent, arguments provide us with finite-size scaling of the fidelity susceptibility at ${\bs\lambda}_c$
\cite{schwandt 09}, \cite{albuquerque 10}:
\begin{equation}
\chi_{\bs e}({\bs\lambda}_c) \sim L^{2/\nu},
\label{susc scaling 1}
\end{equation}
or equivalently, in the small-system regime
\begin{equation}
-\ln {\mathscr{F}}_{\bs e}({\bs\lambda}_c, \delta) \sim \delta^2 L^{2/\nu}.
\label{susc scaling 2}
\end{equation}
Let us note here that in vicinities of some quantum-critical points fidelity oscillates on varying $L$, with an amplitude that is particularly large, close to one, in the small-system regime \cite{rams PRA 11}, \cite{ajk-2}. In such cases, the fidelity susceptibility is not well defined. However, the small-system scaling law (\ref{susc scaling 2}) may still hold but in a generalized sense \cite{ajk-2}. Specifically, it is the envelope of the minima of $-\ln {\mathscr{F}}_{\bs e}({\bs\lambda}_c, \delta)$ that scales according to (\ref{susc scaling 2}).

In the macroscopic-system regime, quantum phase transitions have been studied by means of the so called fidelity per site, a quantity whose logarithm is equal to $N^{-1} \ln {\mathscr{F}}_{\bs e}({\bs\lambda}, \delta)$, where $N=L^D$ is the number of sites in a $D$-dimensional system \cite{barjaktarevic 08}, \cite{zhou 08}.
However, critical scaling of a macroscopic-system fidelity has been considered only very recently by Rams and Damski \cite{rams PRL 11},\cite{rams PRA 11}. These authors have found that, while for small-systems the fidelity scaling is totaly insensitive to the way the critical point ${\bs\lambda}_c$ is approached  by the points ${\bs \lambda} \pm \delta {\bs e}$ (i.e. for instance, whether they are located on one side of the critical point or on the opposite sides), in the case of macroscopic-system the way of approaching the critical point matters.  To make this explicit, Rams and Damski substituted ${\bs\lambda}_c + c \delta {\bs e}$ for ${\bs\lambda}$. By choosing the value of the parameter  $c$, the above mentioned location of  the two points
${\bs\lambda}_c +  c \delta {\bs e} \pm \delta {\bs e}$ with respect to the critical point can be controlled. If $|c| {>} 1$ or  $|c| {<} 1$, then both points are located on one side or on opposite sides of ${\bs\lambda}_c$, respectively. If $|c| {=} 1$, then one of the points coincides with ${\bs\lambda}_c$. Now, the above mentioned independence of the small-system-fidelity scaling on the way the critical point ${\bs\lambda}_c$ is approached  by the points
${\bs \lambda} \pm \delta {\bs e}$ can be expressed as follows:
\begin{equation}
-\ln {\mathscr{F}}_{\bs e}({\bs\lambda}_c + c \delta {\bs e}, \delta) \sim \delta^2 L^{2/\nu}.
\label{susc scaling 3}
\end{equation}
In contrast to the small-system case, the fidelity scaling law for macroscopic systems, derived by Rams and Damski \cite{rams PRL 11}, makes the dependence on parameter $c$ explicit. Provided that the thermodynamic limit of
$N^{-1}\ln {\mathscr{F}}_{\bs e}({\bs\lambda}, \delta)$ does exist, it reads
\begin{equation}
- \ln {\mathscr{F}}_{\bs e}({\bs\lambda}_c + c \delta {\bs e}, \delta) \sim |\delta|^{D\nu} N {\mathscr{A}}_{\bs e}(c),
\label{fid scal}
\end{equation}
where ${\mathscr{A}}_{\bs e}(c)$ is the scaling function.

It should be emphasized that the small-system scaling law (\ref{susc scaling 3}) as well as the macroscopic-system scaling law (\ref{fid scal}) have been derived, using critical-scaling theory arguments, under two conditions. The first one is that there is only one characteristic length scale in the underlying system, which discriminates between small systems and macroscopic systems.
This characteristic length is identified with the effective correlation length
${\tilde{\xi}}_{\bs e}({\bs\lambda}_c + c \delta {\bs e}, \delta) {\equiv} {\tilde{\xi}}_{\bs e}({\bs\lambda}_c, \delta)$.
The second one is that the strict inequality $D \nu {<2}$ holds true \cite{schwandt 09}, \cite{albuquerque 10}, \cite{rams PRL 11}. It is worth to mention that if this condition is satisfied, then  in the small-system regime $\chi_{\bs e}({\bs\lambda}_c)$ or
$-\ln {\mathscr{F}}_{\bs e}({\bs\lambda}_c, \delta)$, formulae (\ref{susc scaling 1}), (\ref{susc scaling 2}), respectively, scale with system's linear size in a superextensive way.

Let $|0 \rangle_{qp}$ and $|\tilde{0} \rangle_{qp}$ be two ground states, the first one for pairs $(\mu,J)$, and the functions $\varepsilon_{\bs k}$, $E_{\bs k}$, the second one for pairs $({\tilde \mu},{\tilde J})$, and the functions ${\tilde \varepsilon_{\bs k}}$, ${\tilde E_{\bs k}}$. As a result of the product structure of the ground states, the quantum fidelity for these states has also a product structure,
\begin{equation}
|_{qp}\langle 0 |\tilde{0} \rangle_{qp}|=
\prod_{\bs k}| \left( u_{\bs k} {\tilde u_{\bs k}} +
|v_{\bs k}| |{\tilde v_{\bs k}}| \exp i (\arg {\tilde v_{\bs k}} - \arg v_{\bs k}) \right) |.
\label{qp_fidelity}
\end{equation}

After using (\ref{uk}), (\ref{vk}) the fidelity of two ground states (\ref{gs}) assumes the form
\begin{equation}
|_{qp}\langle 0 |\tilde{0} \rangle_{qp}|=
\prod_{\bs k} f^{1/2}({\bs k}),\,\, f({\bs k})=
\frac{1}{2}
\left( 1 + \frac{\varepsilon_{\bs k} {\tilde \varepsilon}_{\bs k} + J {\tilde J} (\cos k_1 \pm \cos k_2)^2}{E_{\bs k} {\tilde E}_{\bs k}} \right),
\label{qp_fidelity_2d}
\end{equation}
where the sum of cosine functions has to be taken in the case of the symmetric model and the
difference---in the case of the antisymmetric one.

Let us adapt the general notation introduced in the begining of this section to the considered models. As the location of critical points is uniquely determined  by pairs $(\mu,J)$, we set ${\bs\lambda} {\equiv} (\mu,J)$, hence $|{\bs{\lambda}} \rangle {\equiv} |0\rangle_{qp}$.
Then, in formula (\ref{qp_fidelity_2d}) for fidelity, the functions $\varepsilon_k$, $E_k$, given by (\ref{epsilon}) and (\ref{Ek}), respectively, are calculated at
${\bs\lambda}_c + (c-1) \delta {\bs e}$, while ${\tilde \varepsilon}_k$ and ${\tilde E}_k$ --  at
${\bs\lambda}_c + (c+1) \delta {\bs e}$. Finally, we set
$|_{qp}\langle 0 |\tilde{0} \rangle_{qp}| {\equiv} {\mathscr{F}}_{\bs e}({\bs{\lambda}}_c, \delta)$.

Considering quantum-critical scaling of fidelity, we shall study numerically the sum
\begin{equation}
-\frac{1}{2} \sum_{\bs k} \ln f(\bs k)  \equiv - \ln {\mathscr{F}}_{\bs e}({\bs{\lambda}}_c, \delta),
\label{fid_sum}
\end{equation}
as a function of parameter $\delta$ for fixed system size $N$, or vice versa, in neighborhoods of various critical points, in small- and macroscopic-system regimes. We recall that the values of $k_1$ are obtained with periodic boundary conditions and the values of $k_2$ with antiperiodic ones or vice versa. In all the considered cases the function $\ln f({\bs k})$ is either continuous in the whole square $[0,\pi]^2$
or it has an integrable singularity at some ${\bs k}$ (a discontinuity or a logarithmic divergence). Therefore, in all the considered cases the limit $N {\to} \infty$ of the Riemann sum corresponding to (\ref{fid_sum}) does exist,
\begin{equation}
\lim_{N \to \infty} - N^{-1} \ln {\mathscr{F}}_{\bs e}({\bs{\lambda}}_c, \delta) =
\frac{1}{2\pi^2} \int_{[0,\pi]^2} dk_1 dk_2\,\left(- \ln f(k_1,k_2) \right).
\label{fid_integral}
\end{equation}
Consequently, for given sufficiently small $\delta$ and sufficiently large $N$
\begin{equation}
- \ln {\mathscr{F}}_{\bs e}({\bs{\lambda}}_c, \delta) \approx
\frac{N}{2\pi^2} \int_{[0,\pi]^2} dk_1 dk_2\, \left(-\ln f(k_1,k_2) \right)
\label{fid_integral_N}
\end{equation}
approximately, that is in a macroscopic-system regime $- \ln {\mathscr{F}}_{\bs e}({\bs{\lambda}}_c, \delta)$ scales with the system size as $N$.

Any study of critical scaling involves specifying a critical region, that is a critical point and its neighborhood. The quantum-critical points considered in our paper are displayed in Figs. \ref{diagram2d+} and \ref{diagram2d-}. As for neighborhoods, we have chosen line neighborhoods, each one specified by a unit vector ${\bs e}$ and a range of parameter $\delta$, which are scanned by vectors ${\bs\lambda}_c + (c-1) \delta {\bs e}$ and ${\bs\lambda}_c + (c+1) \delta {\bs e}$ on varying $\delta$. Without any loss of generality only $\delta{>}0$ is considered.

Our aim in the sequel is to  confront the predictions of quantum-critical scaling theory for fidelity with exact results, to find out limitations and advantages of fidelity approach in investigations of quantum-critical points in dimensions $D{>}1$. To the best of our knowledge, this task has never been carried out. We note that the case $D{=}1$ has already been extensively studied in \cite{rams PRL 11}, \cite{rams PRA 11}, \cite{ajk-2} and the results are promising.

We would like to learn to what extent the quantum fidelity is useful for determining the correlation lengths and related universal critical exponents $\nu$ in $D{=}2$ systems, that is in cases where there exists a multitude of those quantities of interest, due to their spatial-direction dependence, and the inequality $D\nu{<}2$ may not be satisfied. Thus, in such cases one or both the basic assumptions of the quantum-critical scaling theory of fidelity are violated.
For this purpose, we consider different kinds of critical points exhibited by the considered systems and calculate
$-\ln {\mathscr{F}}_{\bs e}({\bs{\lambda}}_c, \delta)$ as a function of $\delta$, keeping the linear size $L$ fixed or vice versa.
Then, we make an attempt to determine the intervals of $\delta$ or $L$, where $-\ln {\mathscr{F}}_{\bs e}({\bs{\lambda}}_c, \delta)$ obeys a power law. After that, we try to identify the regime of small system, the one of macroscopic system, and the characteristic length, or lengths, that discriminates between the small- and macroscopic-system regimes -- the location and extent of the crossover regime. In connection with this issue
we note here that according to our results concerning correlation lengths, summarized in sections that follow, for a given linear size of the system,
the effective correlation length increases, we move towards the regime of small system,  by decreasing sufficiently $|\delta|$. In the opposite case
the effective correlation length decreases and we move towards the regime of macroscopic system.
Finally, in regions where $-\ln {\mathscr{F}}_{\bs e}({\bs{\lambda}}_c, \delta)$ obeys a power law, using the scaling laws
(\ref{susc scaling 3}) and (\ref{fid scal}) we read off from the plots of fidelity the values of $\nu$ and compare them with the obtained exact values given in the displayed phase diagrams.

While the two-point correlation function $G({\bs r})$, in particular its large-distance asymptotic behavior, and consequently the defined above effective correlation length ${\tilde{\xi}}_{\bs e}({\bs\lambda}_c, \delta)$ depends on spatial directions,
fidelity ${\mathscr{F}}_{\bs e}({\bs{\lambda}}_c, \delta)$ does not even "know" what a spatial direction is. This fact rises
the interesting question of the relation between the mentioned above characteristic length, or lengths, and a multitude of spatial-direction-dependent, effective correlation lengths ${\tilde{\xi}}_{\bs e}({\bs\lambda}_c, \delta)$. Anticipating our discussion of results presented in the sections that follow, we say only that we have demonstrated that the mentioned multitude of effective correlation lengths is reflected in the existence of more than one crossover regimes whose location is given by the effective correlation lengths in specific spatial directions.

\section{\label{symm} The case of symmetric model}

\begin{figure}
\begin{center}
\includegraphics[width=10cm,clip=on]{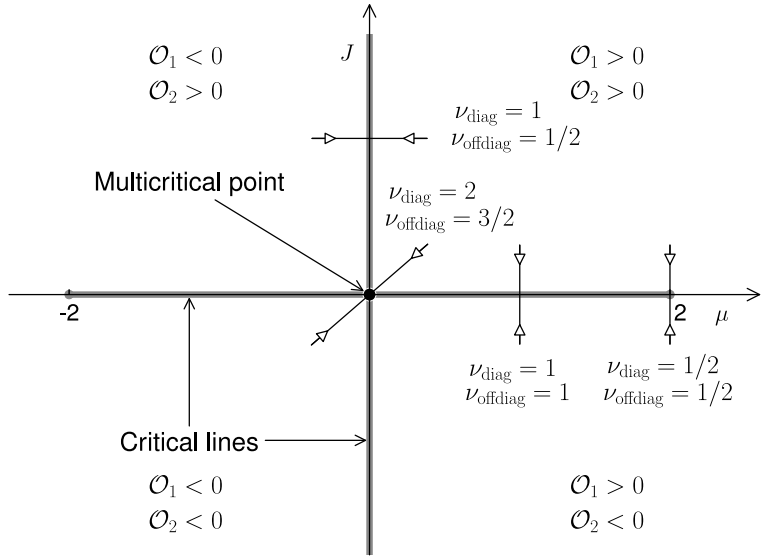}
\caption{\label{diagram2d+} Phase diagram of the symmetric two-dimensional system in the $(\mu,J)$--plane. The set of quantum-critical points consists of the $J$--axis and the closed interval $[-2,2]$ of $\mu$--axis -- thick lines. Those lines constitute also phase boundaries of the four phases, labeled by the order parameters ${\cal{O}}_1$ and ${\cal{O}}_2$. Double arrows indicate the types of neighborhoods of critical points, in which the asymptotic behaviors of $G({\bs r})$ are studied, except neighborhoods of the multicritical point. The universal critical indices $\nu$ in those neighborhoods, whose values are given by the arrows, depend in general on spatial direction, whether it is diagonal ($\nu_{\text{diag}}$) or offdiagonal ($\nu_{\text{offdiag}}$).}
\end{center}
\end{figure}

We can distinguish four ground-state phases labeled by two order parameters, ${\cal{O}}_1$ and ${\cal{O}}_2$, defined as
\begin{equation}
{\cal{O}}_1 = G(0) - \frac{1}{2}, \,\,\,\,{\cal{O}}_2 = - \Delta^{*}h(1,0).
\label{order param}
\end{equation}
To locate the quantum-critical points we solve the equation $E_{\bs k}{=}0$. An inspection of formula (\ref{Ek}) leads immediately to the conclusion that there exist wave vectors ${\bs k}$ such that $E_{k}{=}0$ if and only if $\mu{=}0$ and $J$ is arbitrary (the whole
$J$--axis in the $(\mu,J)$-plane) or $J{=}0$ and $\mu {\in} [-2,2]$ (the closed interval $[-2,2]$ of  the $\mu$--axis).
There are two critical end points $(\pm 2,0)$ and a multicritical point $(0,0)$.
The ground-state phase diagram of the symmetric two-dimensional system is shown in Fig.~\ref{diagram2d+}.
In all the analytic asymptotic formulae presented below, a vicinity of the multicritical point $(0,0)$ is excluded. In particular, for $\mu \to 0$ the $J$-coordinates of  $\mu$-paths have to be away from zero; analogous condition applies to $J$-paths. The case of multicritical point will be discussed separately.

In the stripe $|\mu| {\leq} 2$ of the $(\mu,J)$--plane, but excluding the $\mu{=}0$ and $J{=}0$ lines, the large-distance asymptotic behavior of
$G({\bs r})$ is:
\begin{eqnarray}
{G}(r^{\prime},r^{\prime}) \approx - \textrm{sgn} (\mu)\frac{1}{2\pi} \left( \frac{J^2}{1+J^2} \right)^{1/4}
\frac{\exp(-r^{\prime}/\xi^{(+)})}{r^{\prime}} \cos(\theta r^{\prime}+\phi),
\label{G_diag_asympt_2d+}
\end{eqnarray}
in the diagonal direction, and
\begin{equation}
G(r_1,r_2) \approx
-  \frac{{\cal{C}}_{{\bs r}}}{2\pi} \left( \frac{\mu^2 J^2}{1+J^2} \right)^{1/4} \left( \frac{1+n^2}{n^2} \right)^{1/2}
\frac{\exp \left( -r/\xi_{\text{offdiag}}^{(+)} \right)}{r} \cos (r \theta_{\text{offdiag}} + \phi ),
\label{G offdiag_asympt_dist_2d+}
\end{equation}
with
\begin{eqnarray}
r=r_1\sqrt{(1+n^2)/n^2}, \qquad
{\cal{C}}_{{\bs r}} = \left\{ \begin{array}{ll}
1, & \textrm{if $\mu{>}0$},\\
(-1)^{(r_1+r_2+1)}, & \textrm{if  $\mu{<}0$},
\end{array} \right.
\label{Cr}
\end{eqnarray}
\begin{equation}
\frac{1}{\xi_{\text{offdiag}}^{(+)}} =
\left( \frac{n^2}{1+n^2} \right)^{1/2} \left( \frac{1}{\xi_1} + \frac{1}{n^2} \frac{1}{\xi_2} \right),
\label{xi_offdiag_2d+}
\end{equation}
and,
\begin{equation}
\theta_{\text{offdiag}} = \left( \frac{n^2}{1+n^2} \right)^{1/2} \left( \theta_1 + \frac{1}{n^2} \theta_2 \right),
\label{theta_offdiag_2d+}
\end{equation}
provided that the points $(r_1,r_2)$ become remote from the origin along a ray $r_1/r_2{=} n {=} const$ \cite{ajk-1}.
We recall that by symmetry the large-distance asymptotic behavior of $G({\bs r})$ in offdiagonal directions is the same for
$n {\geq} n_0 {>} 1$  and for $n {\leq} n^{-1}_0 {<} 1$ (specifically we found that one can choose $n_0{=}3$). However, the above formulae
that refer to offdiagonal directions hold only for $n {\geq} n_0 {>} 1$.
The formulae (\ref{G_diag_asympt_2d+}) and (\ref{G offdiag_asympt_dist_2d+}) define the diagonal, $\xi_{\text{diag}}^{(+)}=\sqrt{2}\xi^{(+)}$,
and offdiagonal, $\xi_{\text{offdiag}}^{(+)}$, correlation lengths, respectively.
The parameters $\xi^{(+)}$, $\theta$, $\xi_1$, $\xi_2$, $\theta_1$ and $\theta_2$ will be expressed by $\mu$ and $J$ in vicinities of
critical points in subsections that follow, hence the behavior of $G({\bs r})$ in doubly-asymptotic regions will be specified. Now, we are ready to carry out our programme for specific types of critical points exhibited by the symmetric model.

\subsection{\label{mu0} Critical points at the line $\mu{=}0$}

In this subsection we consider $(\mu,J)$-points approaching along a $\mu$-path a point belonging to any one of the two half lines
of quantum-critical points, given by $\mu{=}0$ and $|J|{\geq} J_0 {>} 0$, for some $J_0$. Then, in terms of $\mu$ and $J$ the parameters determining the large-distance asymptotic behavior of $G({\bs r})$  in (\ref{G_diag_asympt_2d+}) and (\ref{G offdiag_asympt_dist_2d+}) are given by
(see \cite{ajk-1})
\begin{equation}
\frac{1}{\xi^{(+)}} \approx \frac{|\mu J|}{1+J^2}, \qquad
\theta \approx \pi-\frac{|\mu|}{1+J^2},
\label{diag_asympt 2d+ mu=0}
\end{equation}
\begin{equation}
\frac{1}{\xi_1} \approx \sqrt{\frac{2|\mu|}{\sqrt{1+J^2}}}\sin\left(\frac{1}{2}\arctan |J| \right) = - \frac{2}{\xi_2}, \qquad
\theta_1 \approx \pi- 2 \theta_2,
\label{1offdiag_asympt 2d+ mu=0}
\end{equation}
\begin{equation}
\theta_2 \approx \sqrt{\frac{|\mu|}{2\sqrt{1+J^2}}}\cos\left(\frac{1}{2}\arctan |J| \right).
\label{2offdiag_asympt 2d+ mu=0}
\end{equation}

\begin{figure}
\centering
\begin{subfigure}[b]{0.48\textwidth}
\centering
\includegraphics[width=8cm,clip=on]{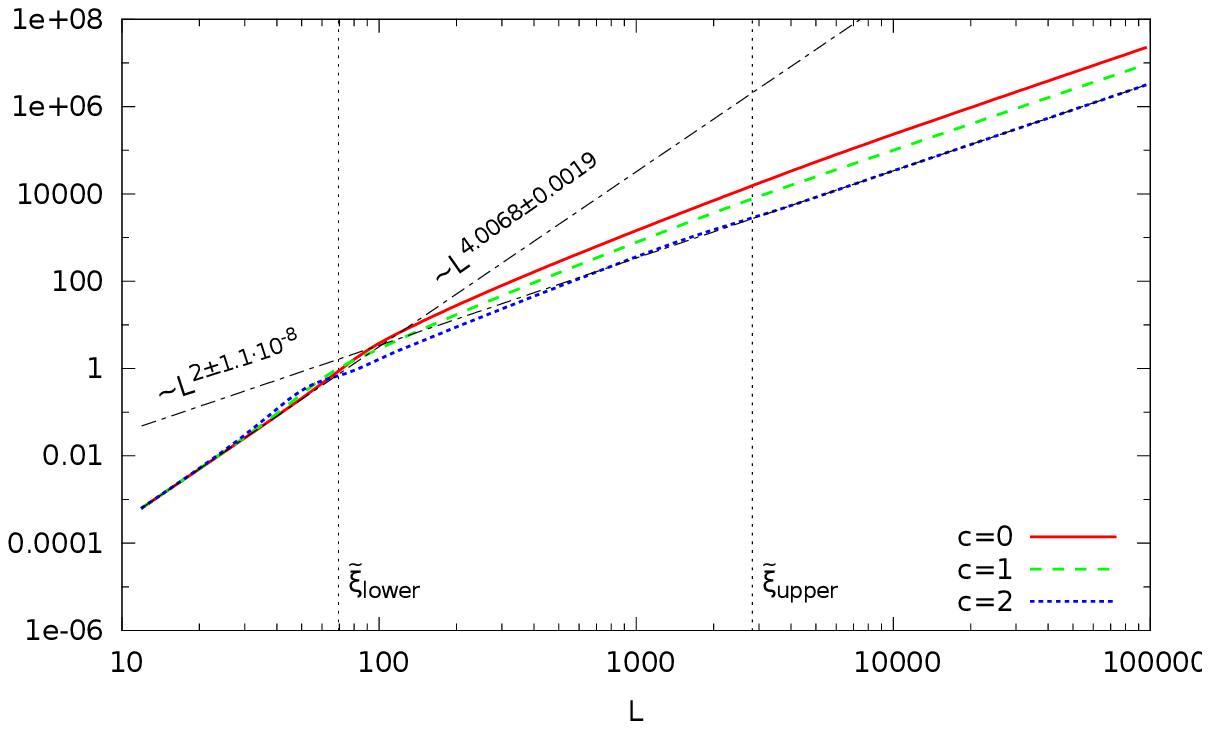}
\end{subfigure}
~
\begin{subfigure}[b]{0.48\textwidth}
\centering
\includegraphics[width=8cm,clip=on]{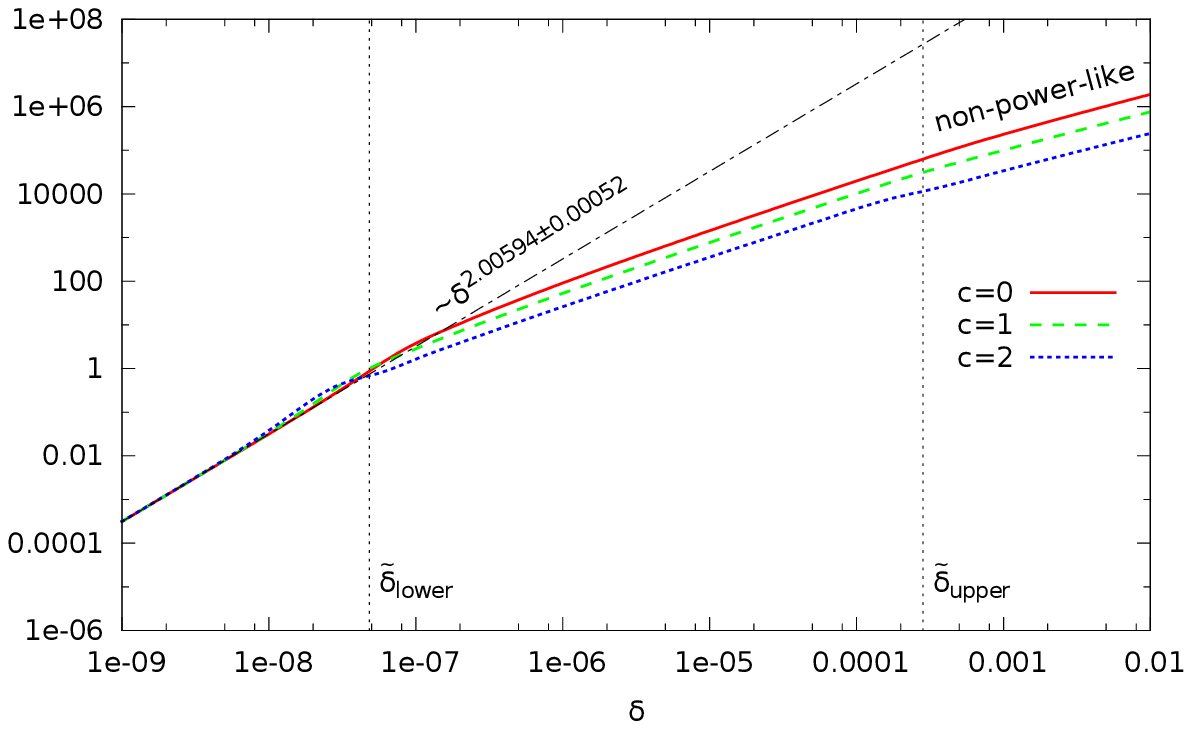}
\end{subfigure}
\caption{\label{2d_lnF_v_L_mu0}\label{2d_lnF_v_ddmu_mu0}(Color online) The symmetric model, critical point $\mu{=}0$ and $J{=}1$,
$\delta{\equiv}\mu$.
Plots of $-\ln {\mathscr{F}}_{(1,0)}((0,1), \delta)$
for three values of $c$: $c{=}0$ -- red line, $c{=}1$ -- green line, $c{=}2$ -- blue line.
Left panel: plots of  $-\ln {\mathscr{F}}_{(1,0)}((0,1), 10^{-3})$ versus $L$.
Right panel: plots of $-\ln {\mathscr{F}}_{(1,0)}((0,1), \delta)$ versus $\delta$ for $L{=}10^4$.
Both plots are in doubly logarithmic scale. Here and in all the figures below, black dashed-dotted straight lines
indicate the power-law scaling. The uncertainties in their slopes have been obtained by the least-square fitting.  }
\end{figure}

In the left and right panels of Fig.~\ref{2d_lnF_v_L_mu0} we show two typical plots of $-\ln {\mathscr{F}}_{(1,0)}((0,1), \delta)$; in the left panel fidelity is plotted versus the system's linear size $L$ while in the right one -- versus the deviation from the critical point  $\delta{\equiv}\mu$. Three regions of different behavior of fidelity, separated by two crossover regions are well visible. Our first goal is to relate the locations of the crossover regions with the correlation lengths of the system. It appears that
in the left panel plot of fidelity, made for fixed $\delta{=}10^{-3}$,  on increasing $L$ the first crossover occurs in vicinity of the effective correlation length ${\tilde{\xi}}_{(1,0)}((0,1), \delta)$ in the axial direction (later denoted ${\tilde{\xi}}_{\text{axial}}(\delta)$),
while the second one -- in vicinity of the effective correlation length ${\tilde{\xi}}_{(1,0)}((0,1), \delta)$ in the diagonal direction (denoted ${\tilde{\xi}}_{\text{diag}}(\delta)$).
With the two effective correlation lengths, ${\tilde{\xi}}_{\text{axial}}(\delta)$ and ${\tilde{\xi}}_{\text{diag}}(\delta)$, we can associate two effective deviations $\tilde{\delta}$, i.e. $\tilde{\delta}_{\text{axial}}(L)$ and $\tilde{\delta}_{\text{diag}}(L)$, defined as the solutions of the equations ${\tilde{\xi}}_{\text{axial}}(\delta){=}L$ and ${\tilde{\xi}}_{\text{diag}}(\delta){=}L$, respectively. It appears that
in the right panel plot of fidelity, made for fixed $L{=}10^4$, on increasing $\delta$ the first crossover occurs in vicinity of the effective deviation $\tilde{\delta}_{\text{axial}}(L)$  while the second one -- in vicinity of the effective deviation $\tilde{\delta}_{\text{diag}}(L)$.
Having identified the locations of the crossover regions with effective correlation lengths in specific directions, we would like to assign a meaning for them, define them using the set off all the spatial direction-dependent correlation lengths.
To this end, let us denote by ${\tilde{\xi}}_{\text{n}}(\delta)$ the effective correlation length ${\tilde{\xi}}_{(1,0)}((0,1), \delta)$ in spatial direction given by $n{>}1$ and let $n_2{>}n_1{>}1$.
From formulae (\ref{xi_offdiag_2d+}), (\ref{diag_asympt 2d+ mu=0}), and (\ref{1offdiag_asympt 2d+ mu=0}) (see also Figs.~15, 16 and comments in section 5 in \cite{ajk-1}), one easily infers  that for sufficiently small $\delta$ the following inequalities hold true:
\begin{equation}
{\tilde{\xi}}_{\text{axial}}(\delta) < {\tilde{\xi}}_{n_2}(\delta) < {\tilde{\xi}}_{n_1}(\delta) < {\tilde{\xi}}_{\text{diag}}(\delta),
\label{tilde xi ineq}
\end{equation}
provided $n_1{>}n_0{>}1$; moreover, ${\tilde{\xi}}_{\text{axial}}(\delta){=}\lim_{n\to\infty}{\tilde{\xi}}_{\text{n}}(\delta)$. However, our numerical results presented in \cite{ajk-1} support the hypothesis that inequalities (\ref{tilde xi ineq}) hold true for any $n{>}1$, and then
${\tilde{\xi}}_{\text{diag}}(\delta){=}\lim_{n\to1}{\tilde{\xi}}_{\text{n}}(\delta)$. Let $\tilde{\delta}_{n}(L)$ be the effective deviation in direction $n$, i.e. the solution of the equation
${\tilde{\xi}}_{\text{n}}(\delta){=}L$ for some sufficiently large $L$.
Since ${\tilde{\xi}}_{\text{n}}(\delta)$ is a decreasing function of $\delta$, inequalities (\ref{tilde xi ineq}) between effective correlation lengths in different directions imply analogous inequalities between the corresponding effective deviations:
\begin{equation}
\tilde{\delta}_{\text{axial}}(L) < \tilde{\delta}_{n_2}(L) < \tilde{\delta}_{n_1}(L) < \tilde{\delta}_{\text{diag}}(L),
\label{tilde xi delta ineq}
\end{equation}
for sufficiently large $L$, where $\tilde{\delta}_{\text{axial}}(L){=}\lim_{n\to\infty}\tilde{\delta}_{n}(L)$, $\delta_{\text{diag}}(L){=}\lim_{n\to 1}\delta_{n}(L)$, and vice versa.
From (\ref{tilde xi ineq}) one concludes that for sufficiently small deviation $\delta$
\begin{equation}
{\tilde{\xi}}_{\text{axial}}(\delta){=}\inf_{n>1}{\tilde{\xi}}_{\text{n}}(\delta) \equiv  {\tilde{\xi}}_{\text{lower}}(\delta) \qquad
\text{and} \qquad {\tilde{\xi}}_{\text{diag}}(\delta){=}\sup_{n>1}{\tilde{\xi}}_{\text{n}}(\delta) \equiv {\tilde{\xi}}_{\text{upper}}(\delta),
\label{tilde xi low upp}
\end{equation}
where we defined the lower effective correlation length ${\tilde{\xi}}_{\text{lower}}(\delta)$ and the upper effective correlation length
${\tilde{\xi}}_{\text{upper}}(\delta)$.
Analogously, for sufficiently large $L$, (\ref{tilde xi delta ineq}) implies that
\begin{equation}
\tilde{\delta}_{\text{axial}}(L){=}\inf_{n>1}\tilde{\delta}_{n}(L) \equiv \tilde{\delta}_{\text{lower}}(L) \qquad
\text{and} \qquad \tilde{\delta}_{\text{diag}}(L){=}\sup_{n>1}\tilde{\delta}_{n}(L) \equiv \tilde{\delta}_{\text{upper}}(L),
\label{delta low upp}
\end{equation}
where we defined the lower effective deviation $\tilde{\delta}_{\text{lower}}(L)$ and the upper effective deviation $\tilde{\delta}_{\text{upper}}(L)$.

Summing up, our numerical results, in particular those displayed in Fig.~\ref{2d_lnF_v_L_mu0}, show that there are two characteristic lengths,
${\tilde{\xi}}_{\text{lower}}(\delta)$ and ${\tilde{\xi}}_{\text{upper}}(\delta)$, or equivalently two characteristic deviations from the critical point, $\tilde{\delta}_{\text{lower}}(L)$ and $\tilde{\delta}_{\text{upper}}(L)$, that mark the crossover regions in the behavior of
$-\ln {\mathscr{F}}_{(1,0)}((0,1), \delta)$ versus $L$ or $\delta$, respectively. Remarkably, in spite of the fact that the fidelity does not depend explicitly on spatial directions $n$, a finger-print of the dependence of correlation properties on spatial directions can be seen in the behavior of the fidelity as a function of $L$ or $\delta$.

We can naturally identify the range of sufficiently large but smaller than ${\tilde{\xi}}_{\text{lower}}(\delta)$ values of $L$, or the range of sufficiently large but smaller than $\tilde{\delta}_{\text{lower}}(L)$ values of $\delta$, as the small-system regime. Then, the ranges of $L$ and $\delta$, given by double inequalities
${\tilde{\xi}}_{\text{lower}}(\delta) {<}L{<} {\tilde{\xi}}_{\text{upper}}(\delta)$ and
$\tilde{\delta}_{\text{lower}}(L) {<}\delta{<} \tilde{\delta}_{\text{upper}}(L)$ can be identified as the mesoscopic-system regimes. Finally, the ranges of $L$ and $\delta$ above ${\tilde{\xi}}_{\text{upper}}(\delta)$ or $\tilde{\delta}_{\text{upper}}(L)$, respectively -- as the macroscopic-system regimes.

Having defined the characteristic lengths, ${\tilde{\xi}}_{\text{lower}}(\delta)$ and ${\tilde{\xi}}_{\text{upper}}(\delta)$, and characteristic deviations from the critical point, $\tilde{\delta}_{\text{lower}}(L)$ and $\tilde{\delta}_{\text{upper}}(L)$, which mark the crossover regions in the behavior of fidelity, and then the regimes of small- and macroscopic-system, we can describe our numerical results, from the perspective of the scaling laws (\ref{susc scaling 3}), (\ref{fid scal}). We note that the divergence of
${\tilde{\xi}}_{\text{lower}}(\delta){=}{\tilde{\xi}}_{\text{axial}}(\delta)$ as $\delta{\to}0$ is characterized by $\nu_{\text{offdiag}}{=}1/2$, which satisfies the condition $D\nu{<}2$. Then, in the small-system regime of $L{<}{\tilde{\xi}}_{\text{lower}}(\delta)$ the $L^4$ scaling of fidelity is consistent with
(\ref{susc scaling 3}) for $\nu{=}1/2$, which matches $\nu_{\text{offdiag}}$. In the macroscopic-system regime of
$L{>}{\tilde{\xi}}_{\text{upper}}(\delta)$ we observe the standard (see section \ref{fidelity}) $L^2$ scaling of fidelity. On the other hand the scaling of fidelity with respect to $\delta$ does not provide any information about the exponent $\nu$. In the small-system regime of $\delta{<}\tilde{\delta}_{\text{lower}}(L)$ we observe the standard (see section \ref{fidelity}) $\delta^2$ scaling of fidelity, while in the macroscopic-system regime of  $\delta{>}\tilde{\delta}_{\text{upper}}(L)$ the fidelity behaves in an anomalous, non-power-law way, which does not allow for estimating $\nu$ via (\ref{fid scal}). We note however, that ${\tilde{\xi}}_{\text{upper}}(\delta)$, which is used to calculate $\tilde{\delta}_{\text{upper}}(L)$, diverges as
$\delta^{-1}$ if $\delta{\to}0$ ($\nu_{\text{diag}}{=}1$), which violates the condition $D\nu{<}2$.

\subsection{\label{0mu2} Critical points in the intervals $J{=}0$, $0{<}|\mu|{<}2$}

In $J$-path neighborhoods of the critical points at the line segments $J{=}0$ and $0{<} |\mu| {<}2$,

\begin{eqnarray}
\frac{1}{\xi^{(+)}} \approx\frac{2|\mu J|}{\sqrt{4-\mu^2}}, \qquad
\theta \approx 2\arccos\frac{|\mu|}{2}.
\label{diag_asympt 2d+ J=0}
\end{eqnarray}
\begin{eqnarray}
\frac{1}{\xi_1} \approx |J|\sqrt{\frac{|\mu|}{2-|\mu|}}, \qquad
\theta_1 \approx \pi - 2\arcsin\sqrt{\frac{|\mu|}{2}} +\frac{1}{2}\sqrt{\frac{|\mu|}{2-|\mu|}}\frac{3-|\mu|}{2-|\mu|}J^2,
\label{1offdiag_asympt 2d+ J=0}
\end{eqnarray}

\begin{eqnarray}
\frac{1}{\xi_2} \approx \frac{|\mu| - 1}{2} \frac{1}{\xi_1}, \qquad
\theta_2 \approx  \frac{1}{2}\sqrt{{|\mu|}{(2-|\mu|)}},
\label{2offdiag_asympt 2d+ J=0}
\end{eqnarray}
for sufficiently small $|J|$ \cite{ajk-1}.
\begin{figure}
\centering
\begin{subfigure}[b]{0.48\textwidth}
\centering
\includegraphics[width=8cm,clip=on]{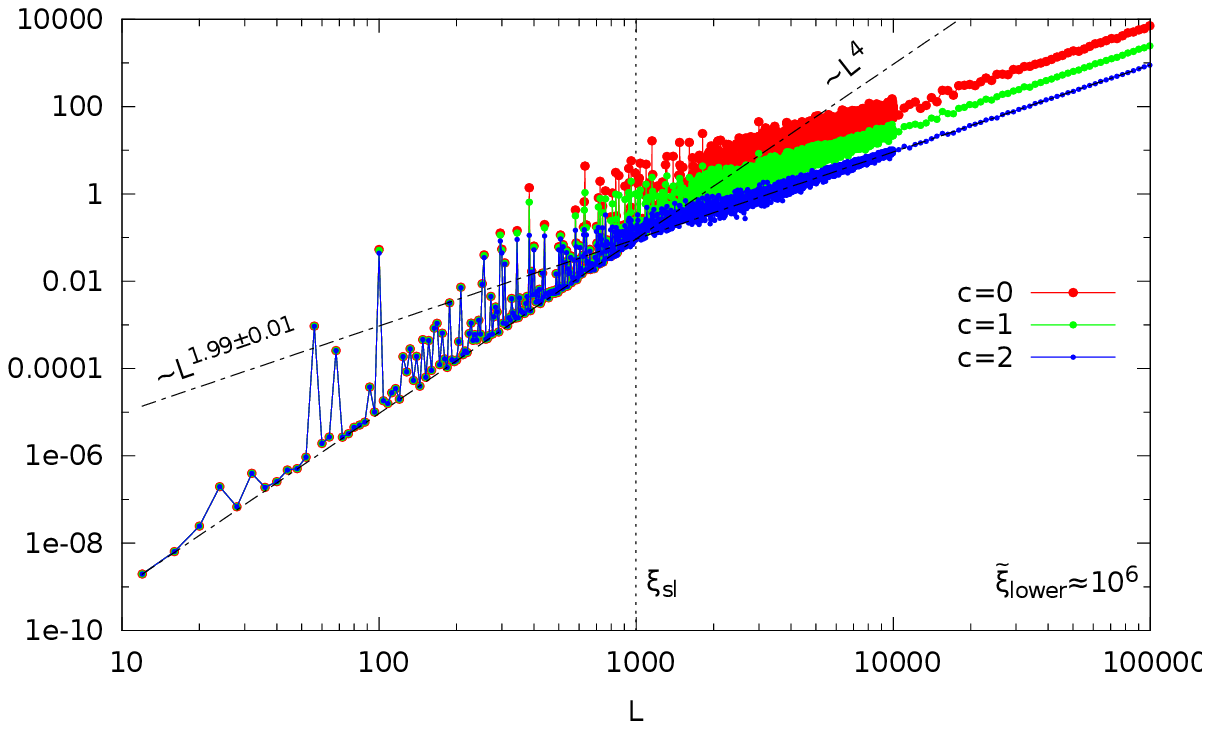}
\end{subfigure}
~
\begin{subfigure}[b]{0.48\textwidth}
\centering
\includegraphics[width=8cm,clip=on]{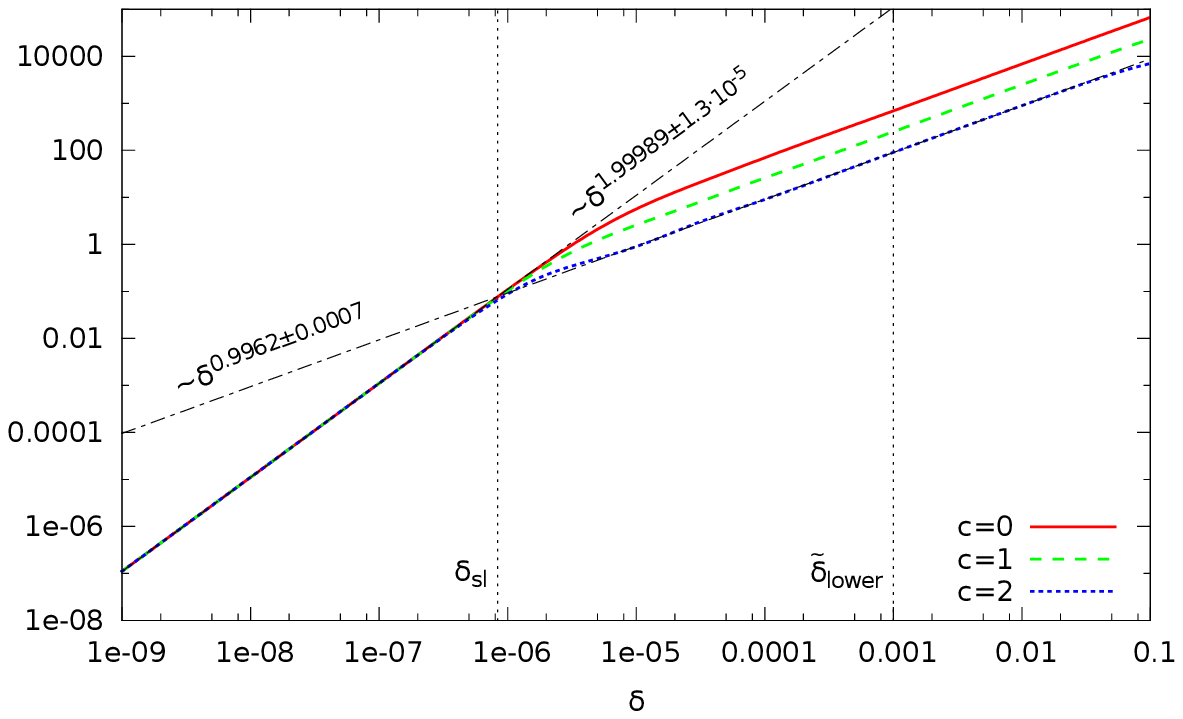}
\end{subfigure}
\caption{\label{2d_lnF_v_L_J0}\label{2d_lnF_v_ddJ_J0}(Color online) The symmetric model, critical point $\mu{=}1$ and $J{=}0$,
$\delta{\equiv}J$.
Plots of $-\ln {\mathscr{F}}_{(0,1)}((1,0), \delta)$
for three values of $c$: $c{=}0$ -- red line, $c{=}1$ -- green line, $c{=}2$ -- blue line.
Left panel: plots of  $-\ln {\mathscr{F}}_{(0,1)}((1,0), 10^{-6})$ versus $L$.
Right panel: plots of $-\ln {\mathscr{F}}_{(0,1)}((1,0), \delta)$ versus $\delta$ for $L=10^3$.
Both plots are in doubly logarithmic scale and the black dashed-dotted straight lines indicate the power-law scaling.
The variable $L$ in the left panel changes by $4$ up to $10\,000$, but
above $10\, 000$ the formula $L\to 4\lceil1.05 L/4\rceil$ is used, where
$\lceil\cdot\rceil$ denotes rounding up to the nearest integer (ceiling
function). The latter way of sampling $L$ reveals the $L^2$ behaviour for sufficiently large $L$,
where for computational as well as presentation reasons, continuing to
change $L$ by $4$ is no longer feasible. This formula has an advantage
of evenly spacing values of $L$ in logarithmic scale but fails to  capture properly the character of the
oscillations for sufficiently small $L$.}
\end{figure}

In those neighborhoods,
the effective correlation lengths ${\tilde{\xi}}(\delta)$, as functions of $\delta {\equiv} J$ and $n$, share the properties that were used in the previous subsection to define the lower and upper effective correlation lengths, ${\tilde{\xi}}_{\text{lower}}(\delta)$ and ${\tilde{\xi}}_{\text{upper}}(\delta)$. Therefore, we can adopt the same definitions of these lengths, the associated effective deviations, $\tilde{\delta}_{\text{lower}}(L)$ and $\tilde{\delta}_{\text{upper}}(L)$, and the small- and macroscopic-system regimes.
In the special case of $\mu{=}1$, all the hierarchy of correlation lengths and the associated deviations collapses,
${\tilde{\xi}}_{\text{lower}}(\delta){\approx}{\tilde{\xi}}_{\text{upper}}(\delta)$ for sufficiently small $\delta$, and
$\tilde{\delta}_{\text{lower}}(L){\approx}\tilde{\delta}_{\text{upper}}(L)$ for sufficiently large $L$.
The critical exponents characterizing the divergence of
${\tilde{\xi}}_{\text{lower}}(\delta)$ and ${\tilde{\xi}}_{\text{upper}}(\delta)$ coincide, $\nu_{\text{offdiag}}{=}\nu_{\text{diag}}{=}1$, and violate the condition $D\nu{<}2$. In Fig.~\ref{2d_lnF_v_L_J0} we show an example of the behavior of fidelity in neighborhoods of the considered critical points. In the left panel, the plot of $-\ln {\mathscr{F}}_{(0,1)}((1,0), 10^{-6})$ versus $L$ exhibits pronounced oscillations, whose amplitude is particularly large for $L{<}\xi_{\text{sl}}$ and decreases with increasing $L$.
We remark that the nature of those oscillations has been revealed in \cite{rams PRA 11}. As we mentioned in section \ref{fidelity} (see also \cite{ajk-2}) on presenting small-system scaling law,
in case of oscillating fidelity one should consider scaling in the generalized sense. That is, in such a case it is the envelope of the minima of
$-\ln {\mathscr{F}}_{\bs e}({\bs{\lambda}}_c, \delta)$ that is the right quantity whose scaling should be studied. In the left panel,
one crossover region separating different power-law behaviors, in the generalized sense, marked by some $\xi_{\text{sl}}(\delta)$, is visible.
However, $\xi_{\text{sl}}$ does not match ${\tilde{\xi}}_{\text{lower}}(\delta){\approx}{\tilde{\xi}}_{\text{upper}}(\delta)$, whose values are a lot larger than $\xi_{\text{sl}}(\delta)$.
If we naively identify the regime of $L{<}\xi_{\text{sl}}$ as the small-system regime, then the observed $L^4$-scaling of fidelity implies, via (\ref{susc scaling 3}), $\nu{=}1/2$, which does not match the exact value $1$ of this exponent. Moreover, $L^4$-scaling of
$-\ln {\mathscr{F}}_{\bs e}({\bs{\lambda}}_c, \delta)$, consequently superextensivity, interestingly occurs despite the fact that the condition of superextensive behavior, which follows from critical-scaling theory, $D\nu{<}2$, is violated.
For $L{>}\xi_{\text{sl}}$, identified as the macroscopic-system regime we observe the standard
$L^2$-scaling. In the right panel, the plot of $-\ln {\mathscr{F}}_{(0,1)}((1,0), \delta)$ versus $\delta$  exhibits also one crossover region, separating different power-law behaviors, marked by some $\delta_{\text{sl}}$, which is the solution of the equation $\xi_{\text{sl}}(\delta){=}L$.
The power law (\ref{fid scal}) applied for $\delta$ well above $\delta_{\text{sl}}$ gives the incorrect result $\nu{=}1/2$. We note that similar results hold for other values of $\mu$, $|\mu|{\neq}1$. The only difference is that ${\tilde{\xi}}_{\text{lower}}(\delta){<}{\tilde{\xi}}_{\text{upper}}(\delta)$, but still ${\tilde{\xi}}_{\text{lower}}(\delta){\gg}\xi_{\text{sl}}(\delta)$; thus, ${\tilde{\xi}}_{\text{lower}}(\delta)$ does not locate the crossover region correctly.

\subsection{\label{mu2} The end critical points $J{=}0$, $|\mu|{=}2$}

If $(\mu,J)$-points approach along the $J$-path one of the two end critical points, $|\mu|{=}2$ and $J{=}0$,
\begin{equation}
\frac{1}{\xi^{(+)}} \approx 2|J|^{1/2}, \qquad
\theta \approx2|J|^{1/2},
\label{diag_asympt 2d+ mu=2}
\end{equation}
\begin{equation}
\frac{1}{\xi_1} \approx \sqrt{2|J|} = \frac{2}{\xi_2}, \qquad
\theta_1 \approx \sqrt{2|J|} = 2\theta_2,
\label{offdiag_asympt 2d+ mu=2}
\end{equation}
for sufficiently small $J$ \cite{ajk-1}.

\begin{figure}
\centering
\begin{subfigure}[b]{0.48\textwidth}
\centering
\includegraphics[width=8cm,clip=on]{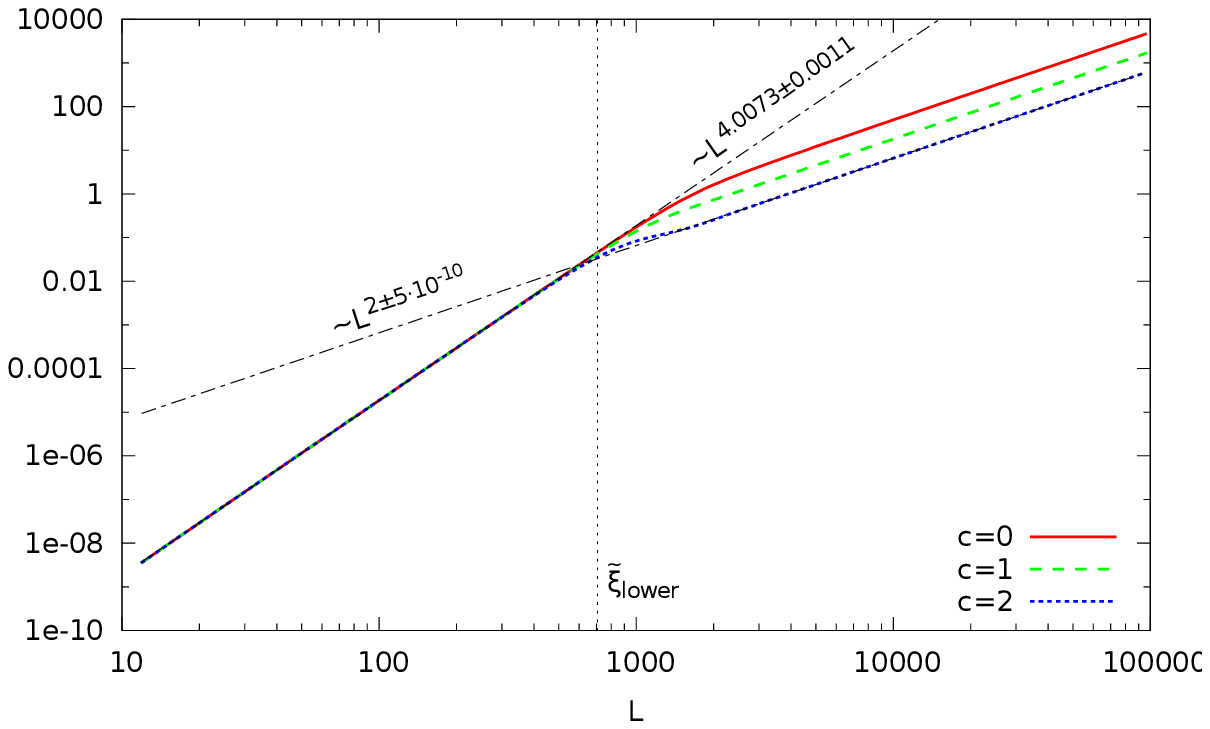}
\end{subfigure}
~
\begin{subfigure}[b]{0.48\textwidth}
\centering
\includegraphics[width=8cm,clip=on]{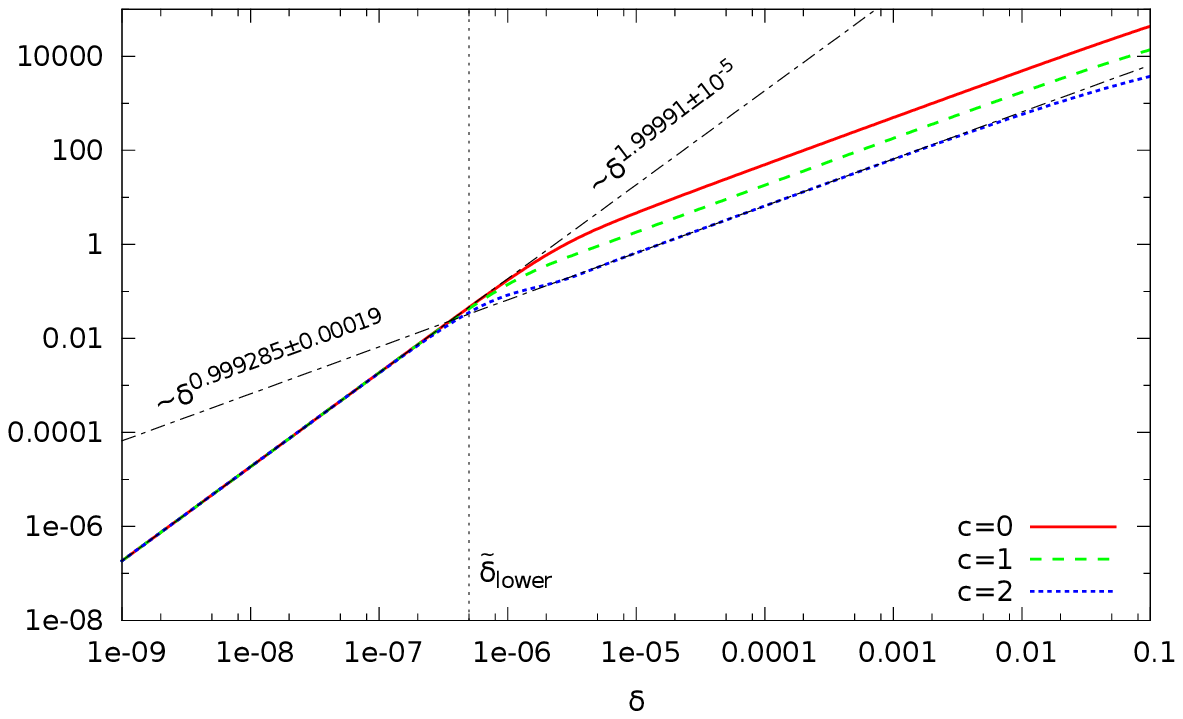}
\end{subfigure}
\caption{\label{2d_lnF_v_L_J0ep}\label{2d_lnF_v_ddJ_J0ep}(Color online) The symmetric model, critical point $\mu{=}2$ and $J{=}0$,
$\delta{\equiv}J$.
Plots of $-\ln {\mathscr{F}}_{(0,1)}((2,0), \delta)$
for three values of $c$: $c{=}0$ -- red line, $c{=}1$ -- green line, $c{=}2$ -- blue line.
Left panel: plots of  $-\ln {\mathscr{F}}_{(0,1)}((2,0), 10^{-6})$ versus $L$.
Right panel: plots of $-\ln {\mathscr{F}}_{(0,1)}((2,0), \delta)$ versus $\delta$ for $L=10^3$.
Both plots are in doubly logarithmic scale and the black dashed-dotted straight lines indicate the power-law scaling. For more details see the text.}
\end{figure}

In a $J$-path vicinity of one of the end critical points $J{=}0$ and $|\mu|{=}2$, setting $\delta{\equiv}J$,  we can adopt the same definitions of ${\tilde{\xi}}_{\text{lower}}(\delta)$ and ${\tilde{\xi}}_{\text{upper}}(\delta)$ of the small-, mesoscopic- and macroscopic-system regimes as in subsections \ref{mu0} and \ref{0mu2}.
However, using (\ref{diag_asympt 2d+ mu=2}) and (\ref{offdiag_asympt 2d+ mu=2}) one verifies easily that, for sufficiently small $J {\equiv} \delta$,
${\tilde{\xi}}_{\text{axial}}(\delta)\approx {\tilde{\xi}}_{\text{n}}(\delta)\approx{\tilde{\xi}}_{\text{diag}}(\delta)$, hence
${\tilde{\xi}}_{\text{lower}}(\delta)\approx {\tilde{\xi}}_{\text{upper}}(\delta)$.
Therefore, there is no mezoscopic-system regime, there is only one characteristic length, say ${\tilde{\xi}}_{\text{lower}}(\delta)$, one corresponding $\tilde{\delta}_{\text{lower}}(L)$, that discriminates between the small- and macroscopic-system regimes. Indeed, in Fig.~\ref{2d_lnF_v_L_J0ep},
in the left panel, where fidelity is plotted against $L$, as well as in the right panel, where fidelity is plotted against $\delta$, only one crossover region, marked by ${\tilde{\xi}}_{\text{lower}}(\delta)$ or $\tilde{\delta}_{\text{lower}}(L)$, respectively, is visible.
The critical exponents characterizing the divergence of
${\tilde{\xi}}_{\text{lower}}(\delta)$ and ${\tilde{\xi}}_{\text{upper}}(\delta)$ coincide, $\nu_{\text{offdiag}}{=}\nu_{\text{diag}}{=}1/2$, and satisfy the condition $D\nu{<}2$.
In the small-system regime, $L{<}{\tilde{\xi}}_{\text{lower}}(\delta)$ (left panel), fidelity scales as $L^4$, which via (\ref{susc scaling 3}) gives the correct value of the exponent $\nu{=}1/2$, while in the macroscopic-system regime, $L{>}{\tilde{\xi}}_{\text{lower}}(\delta)$, the standard
$L^2$-scaling is observed. Then, in the small-system regime, $\delta{<}\tilde{\delta}_{\text{lower}}(L)$, fidelity exhibits the standard
$\delta^2$-scaling, while in the macroscopic-system regime, $\delta{>}\tilde{\delta}_{\text{lower}}(L)$, it scales as $\delta$, which via
(\ref{fid scal}) gives again the correct exponent $\nu{=}1/2$.

\subsection{\label{mu0J0} The multicritical point $\mu{=}0{=}J$}

\begin{figure}
\centering
\begin{subfigure}[b]{0.48\textwidth}
\centering
\includegraphics[width=8cm,clip=on]{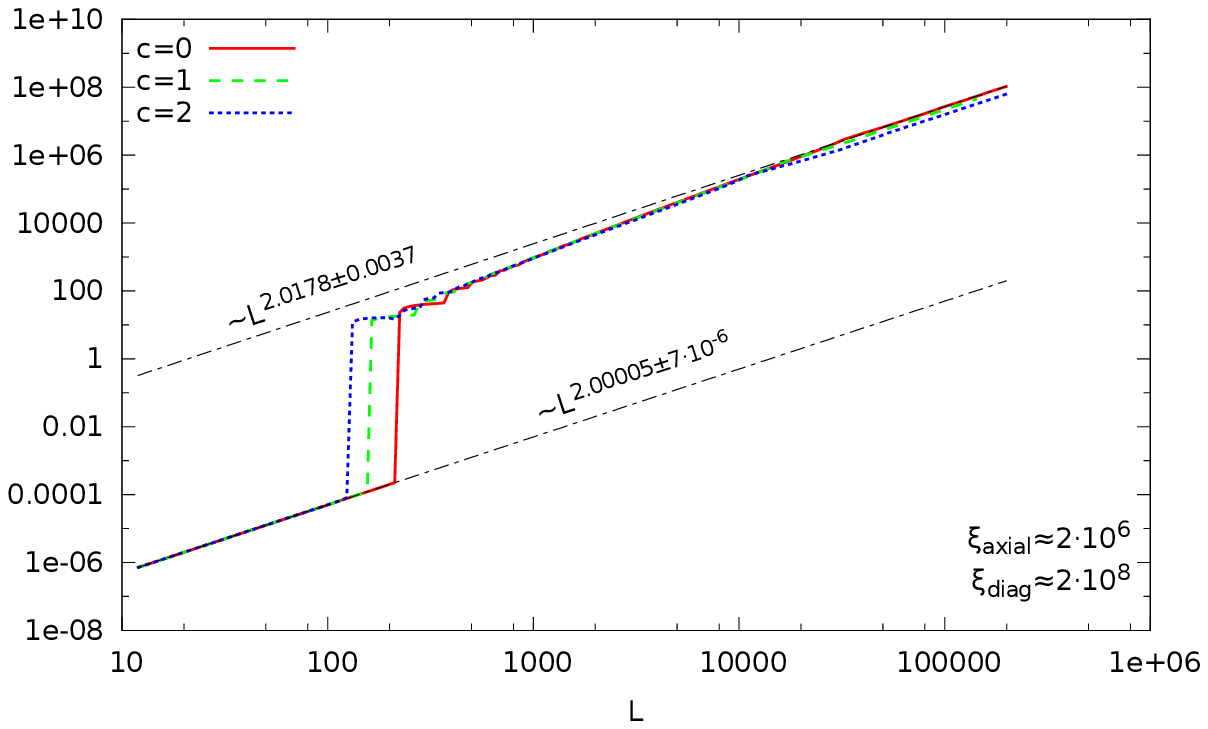}
\end{subfigure}
~
\begin{subfigure}[b]{0.48\textwidth}
\centering
\includegraphics[width=8cm,clip=on]{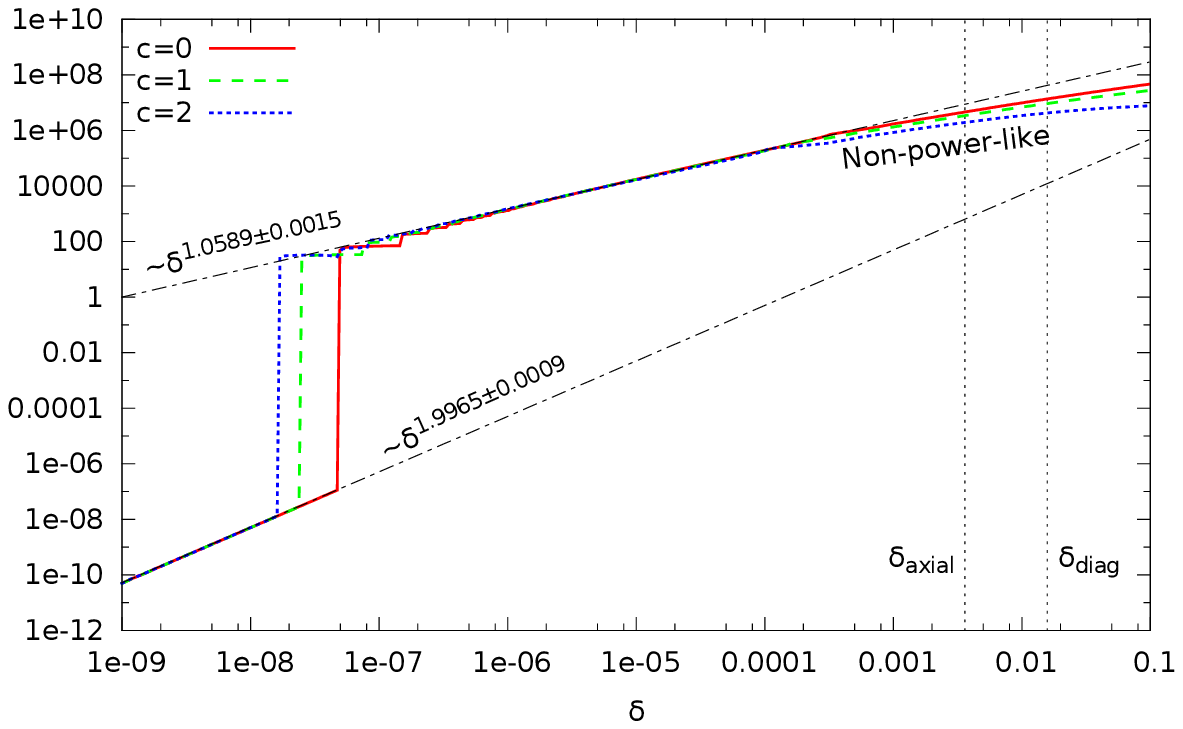}
\end{subfigure}
\caption{\label{2d_lnF_v_L_mcp}\label{2d_lnF_v_d_mcp}(Color online) The symmetric model, critical point $\mu{=}J{=}0$, $\delta{\equiv}\mu{=}J$.
Plots of $-\ln {\mathscr{F}}_{(1,1)}((0,0), \delta)$
for three values of $c$: $c{=}0$ -- red line, $c{=}1$ -- green line, $c{=}2$ -- blue line.
Left panel: plots of  $-\ln {\mathscr{F}}_{(1,1)}((0,0), 10^{-4})$ versus $L$.
The calculated effective correlation lengths ${\tilde{\xi}}_{\text{axial}}(\delta)$ and ${\tilde{\xi}}_{\text{diag}}(\delta)$ are too large to be depicted in the figure.
Right panel: plots of $-\ln {\mathscr{F}}_{(1,1)}((0,0), \delta)$ versus $\delta$ for $L=10^4$.
The effective deviations ${\tilde{\delta}}_{\text{axial}}(L)$ and ${\tilde{\delta}}_{\text{diag}}(L)$ are well above $\delta_{mm}$.
Both plots are in doubly logarithmic scale and the black dashed-dotted straight lines indicate the power-law scaling. For more details see the text.}
\end{figure}

In this case we cannot provide any analytic results concerning the correlation lengths and associated critical indices shown in the phase digram of Fig.~\ref{diagram2d+}. These quantities have been obtained numerically by studying the behavior of $G({\bs r})$ in doubly asymptotic regions, where the multicritical point has been approached along the $45^{\circ}$-path ($\mu{=}J$). To determine $\nu_{\text{offdiag}}$, a few spatial directions, including the axial ones, have been taken into account.
Both the exponents extracted from the asymptotic behavior of $G({\bs r})$, i.e. $\nu_{\text{offdiag}}{=}3/2$ and $\nu_{\text{diag}}{=}2$, violate the condition $D\nu{<}2$. To the best of our knowledge, $\mu{=}0{=}J$ quantum-critical point is the first instance of quantum critical point whose associated indices $\nu$ are so large that they satisfy the strict inequality $D\nu{>}2$.

The plots of fidelity are shown in Fig.~\ref{2d_lnF_v_L_mcp}, against $L$ -- in the left panel and against $\delta{\equiv}\mu{=}J$ -- in the right one. Two crossover regions, marked by $\xi_{sm}$, $\xi_{mm}$, with $\xi_{sm} {<} \xi_{mm}$, and $\delta_{sm}$, $\delta_{mm}$, with $\delta_{sm} {<} \delta_{mm}$, respectively, are visible. Numerically calculated effective correlation lengths
${\tilde{\xi}}_{\text{axial}}(\delta)$ and ${\tilde{\xi}}_{\text{diag}}(\delta)$ are so much larger than $\xi_{mm}$ that they cannot be displayed in the left panel of Fig.~\ref{2d_lnF_v_L_mcp}. The corresponding ${\tilde{\delta}}_{\text{axial}}(L)$ and ${\tilde{\delta}}_{\text{diag}}(L)$ are larger than $\delta_{mm}$ as well but they are visible in the right panel.
If we naively, ignoring critical scaling theory, identify the region of
$L{<}\xi_{sm}$ ($\delta{<}\delta_{sm}$) as small-system regime and that of $L{>}\xi_{mm}$ ($\delta{>}\delta_{mm}$) -- as macroscopic-system regime, we can describe them from the perspective of the fidelity scaling laws. We observe the standard $L^2$-scaling of fidelity for $L{>}\xi_{mm}$  and
the standard $\delta^2$-scaling for $\delta{<}\delta_{sm}$. For $\delta{>}\delta_{mm}$ no prediction for $\nu$ can be made, since fidelity scales with
$\delta$ in a non-power-law way. For $L{<}\xi_{sm}$, fidelity scales as $L^2$, which gives, via (\ref{susc scaling 3}), $\nu{=}1$ -- the value well below any one of the calculated critical exponents.
However, the extensive scaling of fidelity for $L{<}\xi_{sm}$ suggests that there may exist a subextensive contribution to fidelity, which is not visible at the first sight. To this end we take a closer look at the scaling of fidelity in a linear neighborhood of the multicritical point, for systems whose linear size $L$ is small in the sense that $\delta L{\ll}1$. Specifically, we consider a neighborhood of $(0,0)$ quantum-critical point given as follows:
$\mu{=}(c-1)\delta$, $\tilde{\mu}{=}(c+1)\delta$, $J{=}(c-1)\delta$, $\tilde{J}{=}(c+1)\delta$. In this neighborhood, the function $f(\bs k)$, given in (\ref{qp_fidelity_2d}), assumes the form
\begin{equation}
f(\bs k) = \frac{1}{2} + \frac{1}{2}\frac
{[(c+1)\delta-p][(c-1)\delta-p] + (c^2-1)\delta^2p^2}
{\{(c+1)^2\delta^2p^2 + [(c+1)\delta-p]^2 \}^{1/2} \{(c-1)^2\delta^2p^2 +  [(c-1)\delta-p]^2 \}^{1/2}},
\label{f mcp}
\end{equation}
where $p{=}\cos k_1 + \cos k_2$, which after expanding to the fourth order in $\delta$ gives
\begin{equation}
-\frac{1}{2}\ln f(\bs k) \approx \frac{1}{2}\delta^2 + \frac{2c}{p}\delta^3 + \left( \frac{5c^2 + 1}{p^2} - c^2 -\frac{1}{4} \right)\delta^4.
\label{f mcp delta}
\end{equation}
Summing the expansion (\ref{f mcp delta}) over the Brillouin zone corresponding to periodic boundary conditions in one direction and antiperiodic ones in the orthogonal direction, with $L$ even, and using the identities: $\sum_{\bs k} 1/p{=}0$ and $\sum_{\bs k} 1/p^2{=}L^4/4$,  we obtain the following expansion of fidelity with respect to $\delta L$: for $\delta L {\ll} 1$
\begin{equation}
-4\ln {\mathscr{F}}_{(1,1)}((0,0), \delta) \approx a(\delta L)^2 + b (\delta L)^4,
\label{lnF mcp deltaL}
\end{equation}
where
\begin{equation}
a = 2 - \left(4 c^2 +  1 \right)\delta^2,\,\,\,\,b = 5c^2 + 1.
\label{a,b}
\end{equation}

\begin{figure}
\centering
\begin{subfigure}[b]{0.48\textwidth}
\centering
\includegraphics[width=8cm,clip=on]{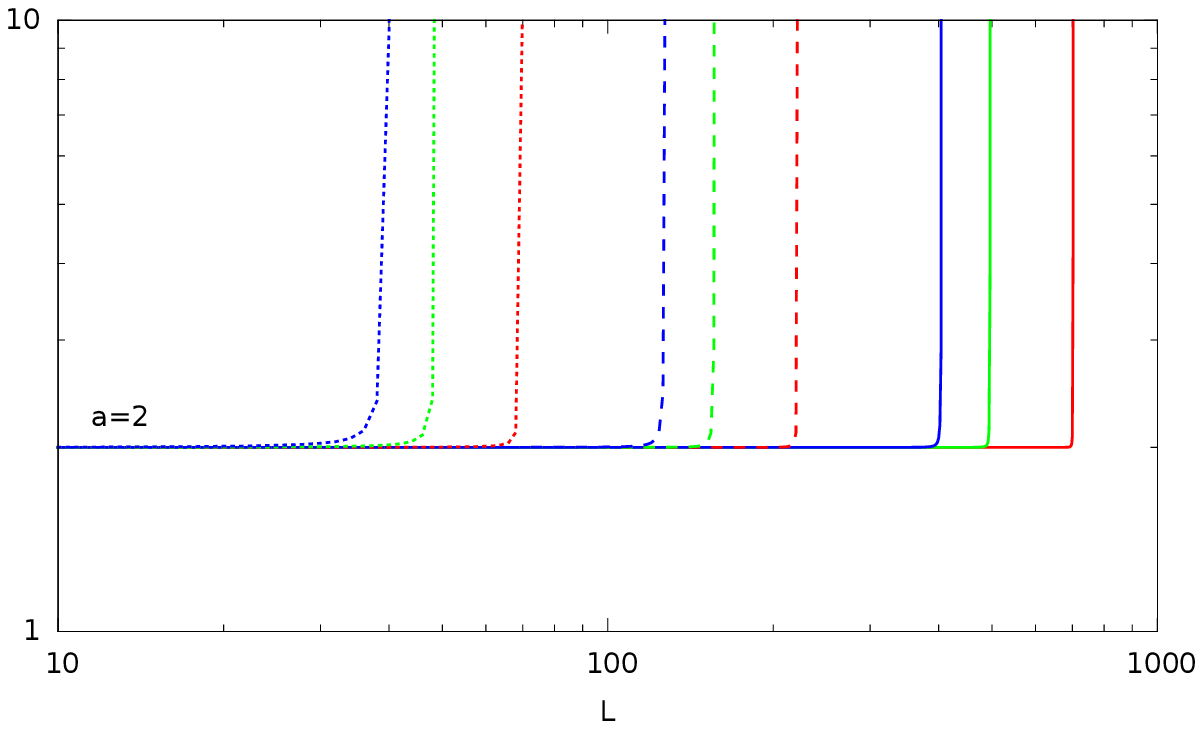}
\end{subfigure}
~
\begin{subfigure}[b]{0.48\textwidth}
\centering
\includegraphics[width=8cm,clip=on]{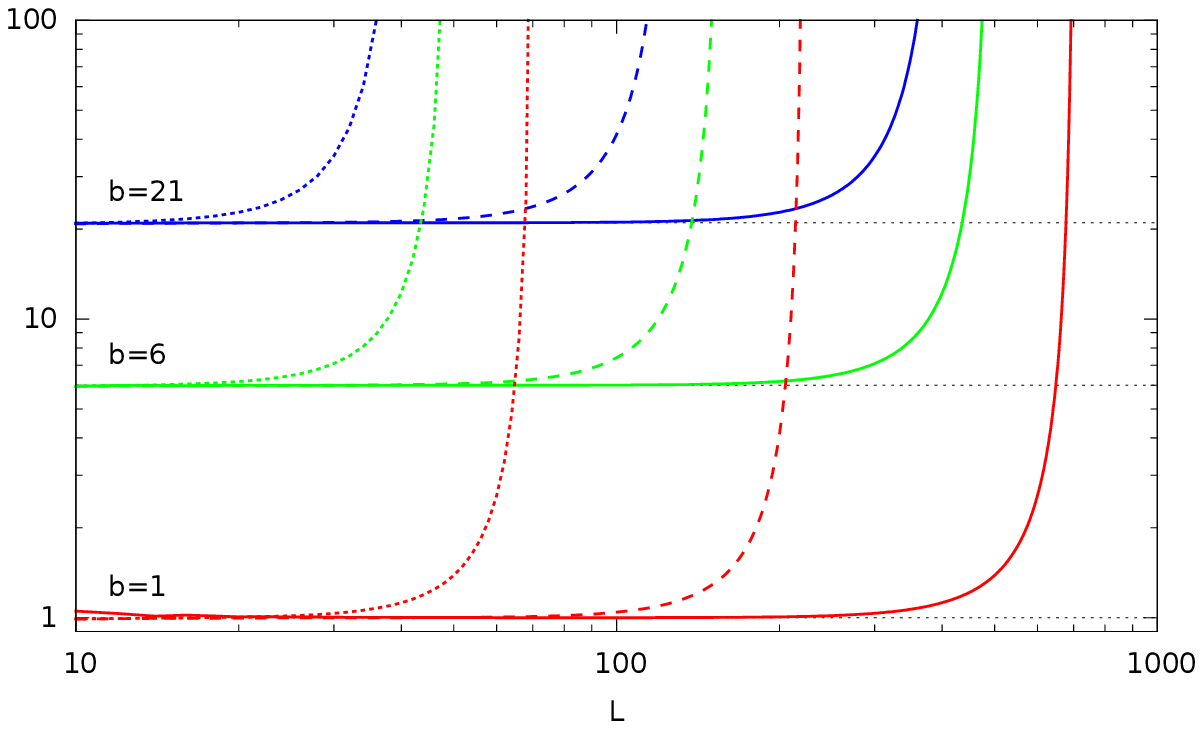}
\end{subfigure}
\caption{\label{2d_mcp_subleading_a4}\label{2d_mcp_subleading_b4}(Color online) The symmetric model, critical point $\mu{=}J{=}0$, $\delta{\equiv}\mu{=}J$.
Numerical determination of the coefficients $a$ and $b$ in the expansion (\ref{lnF mcp deltaL}).
In each panel there are 6 plots,
for three values of $c$: $c{=}0$ -- red line, $c{=}1$ -- green line, $c{=}2$ -- blue line, and for each $c$ for three values of $\delta$, from left to right: $\delta{=}10^{-3}$--dotted line, $\delta{=}10^{-4}$--dashed line, $\delta{=}10^{-5}$--continuous line.
Left panel: plots of  $-4\ln {\mathscr{F}}_{(1,1)}((0,0), \delta)/(\delta L)^2 \approx a + b(\delta L)^2$ versus $L$.
Right panel: plots of $\left( -4\ln {\mathscr{F}}_{(1,1)}((0,0), \delta)/(\delta L)^2 - a \right)/(\delta L)^2 \approx b$ versus $L$.
All the plots are in doubly logarithmic scale. }
\end{figure}
The expansion (\ref{lnF mcp deltaL}) is illustrated in Fig.~\ref{2d_mcp_subleading_a4}, where the values of the coefficients $a$ and $b$ are determined numerically; it is well visible that both the coefficients are approximately constant in systems whose linear size satisfies the inequality $\delta L {\ll} 1$. Thus, there is no subextensive correction to the extensive term; the lowest order correction to the first term, proportional to $L^2$, is superextensive, proportional to $L^4$.

Let us note here that the critical indices characterizing multicritical points of some 1D models satisfy the condition $D\nu{<}2$ but analogous analysis of the behavior of fidelity shows that the laws of quantum-critical scaling of fidelity are violated and one is not able to determine correct values of critical index $\nu$ (see \cite{rams PRA 11}, \cite{ajk-2}). Summarizing, it appears that multicritical points are not amenable to the kind of scaling analysis that we try to perform.

\section{\label{antisymm} The case of antisymmetric model}

Let us note that by applying a mean-field approximation to
\begin{equation}
 \sum_{{\bs k},\sigma}\varepsilon_{\bs k} c^{\dagger}_{{\bs k},\sigma} c_{{\bs k},\sigma}
 + J \sum_{{\bs l},i} {\bs S_{{\bs l}} } {\bs S_{{\bs l}+{\bs e}_i} },
\label{spin ham}
\end{equation}
where ${\bs S_{{\bs l}}}$ stands for the spin operator of a spin $1/2$ fermion at site ${\bs l}$ of the underlying lattice (for details of the notation see section \ref{models}) one obtains the Hamiltonian of the antisymmetric model but together with an equation relating its parameters. In distinction to our set up, the parameters $\Delta_{i}$ of the mean-field Hamiltonian are no longer free parameters but are given implicitly by those solutions of the equations
\begin{equation}
\Delta_{i} = - \langle  a_{{\bs l},\uparrow} a_{{\bs l}+{\bs e}_i,\downarrow} -
a_{{\bs l},\downarrow} a_{{\bs l}+{\bs e}_i,\uparrow} \rangle,
\label{delta}
\end{equation}
that minimize the ground-state energy---physical solutions in the context of system given by (\ref{spin ham}),
where the brackets denote the Gibbs average calculated with the mean-field Hamiltonian.
When the underlying lattice is a square lattice, it turns out that the physical solutions satisfy the condition $\Delta_1{=}-\Delta_2$, which corresponds to the so called $d_{x^2-y^2}$ pairing in theory of $d$-wave superconductivity.

\begin{figure}
\begin{center}
\includegraphics[width=10cm,clip=on]{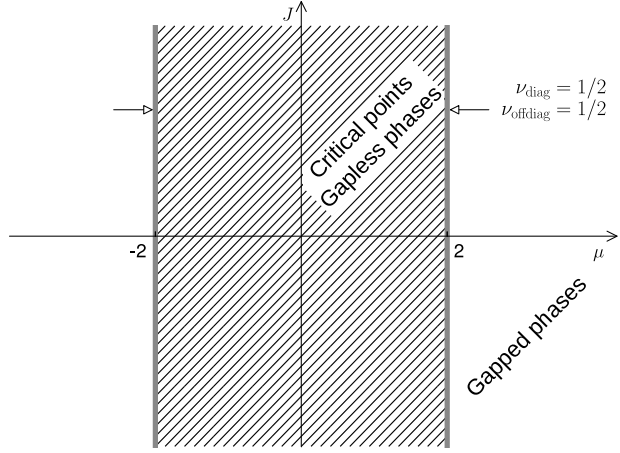}
\caption{\label{diagram2d-} Phase diagram of the antisymmetric two-dimensional system in the $(\mu,J)$-plane. The set of critical points constitutes the stripe  $|\mu|{\leq} 2$.}
\end{center}
\end{figure}

In distinction to the previously considered case of the symmetric two-dimensional model, where the critical points are located at straight, intersecting  lines, the quantum-critical points of the antisymmetric two-dimensional model fill up the stripe that extends between the two lines $|\mu|{=}2$, see the phase diagram in Fig.~\ref{diagram2d-}.
Therefore, there are only two  doubly-asymptotic regions of interest, where analytic formulae for $G({\bs r})$ can be derived, namely those where $(\mu,J)$-points, with $|\mu|{>}2$ and $|J|$ not too close to zero, approach along a $\mu$-path a point  belonging to one of the lines $|\mu|{=}2$. In these regions the asymptotic behavior of $G({\bs r})$
in the diagonal direction is
\begin{eqnarray}
{G}(r^{\prime},r^{\prime}) \approx - \textrm{sgn} (\mu) \frac{J}{4\pi \xi^{(-)}} \frac{\exp(-r^{\prime}/\xi^{(-)})}{{r^{\prime}}},
\label{G diag_asympt 2d-}
\end{eqnarray}
where
\begin{equation}
\frac{1}{\xi^{(-)}} \approx 2\sqrt{\frac{|\mu| - 2}{1+J^2}},
\label{xi diag_2d-}
\end{equation}
that is, in the considered doubly-asymptotic region, the correlation length in the diagonal direction amounts to
$\xi^{(-)}_{\text{diag}} {=} \sqrt{2}\xi^{(-)}$ \cite{ajk-1}.
Then, in offdiagonal directions we obtained the asymptotic formula
\begin{equation}
G(r_1,r_2) \approx
-  \frac{{\cal{C}}_{{\bs r}}}{2\pi} \left( \frac{\mu^2 J^2}{1+J^2} \right)^{1/4} \left( \frac{1+n^2}{n^2} \right)^{1/2}
\frac{\exp \left( -r/\xi_{\text{offdiag}}^{(-)} \right)}{r} \cos (r \theta^{(-)}_{\text{offdiag}} + \phi^{(-)} ),
\label{G offdiag_asympt_dist_2d-}
\end{equation}
with
\begin{equation}
\frac{1}{\xi_{\text{offdiag}}^{(-)}} =
\left( \frac{n^2}{1+n^2} \right)^{1/2} \left( \frac{1}{\xi^{(-)}_1} + \frac{1}{n^2} \frac{1}{\xi^{(-)}_2} \right),
\label{xi_offdiag_2d-}
\end{equation}
and
\begin{equation}
\theta^{(-)}_{\text{offdiag}} = \left( \frac{n^2}{1+n^2} \right)^{1/2} \left( \theta^{(-)}_1 + \frac{1}{n^2} \theta^{(-)}_2 \right),
\label{theta_offdiag_2d-}
\end{equation}
provided that the points $(r_1,r_2)$ become remote from the origin along a ray $r_1/r_2{=} n {=} const$ \cite{ajk-1}.
Sufficiently close to the lines $|\mu|{=}2$, the values of $\xi_1^{(-)}$, $\xi_2^{(-)}$, $\theta_1^{(-)}$, $\theta_2^{(-)}$, and $\phi^{(-)}$, which determine the offdiagonal correlation length $\xi_{\text{offdiag}}^{(-)}$ in a direction $n$ are given as follows (see \cite{ajk-1}):
\begin{equation}
\frac{1}{\xi_1^{(-)}} \approx \sqrt{\frac{\sqrt{1+J^2}+1}{1+J^2}}\sqrt{|\mu| - 2}, \qquad
\theta_1^{(-)} \approx \sqrt{\frac{\sqrt{1+J^2}-1}{1+J^2}}\sqrt{|\mu| - 2}, \qquad
\phi^{(-)} = \frac{\pi}{4},
\label{xi1 theta1 phi 2d-}
\end{equation}
\begin{equation}
\frac{1}{\xi_2^{(-)}} \approx -\left( \frac{1}{2} - \frac{1}{\sqrt{1+J^2}} \right) \frac{1}{\xi_1^{(-)}},
\label{xi2 2d-}
\end{equation}
\begin{equation}
\theta_2^{(-)} \approx -\left( \frac{1}{2} + \frac{1}{\sqrt{1+J^2}} \right) \theta_1^{(-)}.
\label{theta2 2d-}
\end{equation}

\begin{figure}
\centering
\begin{subfigure}[b]{0.48\textwidth}
\centering
\includegraphics[width=8cm,clip=on]{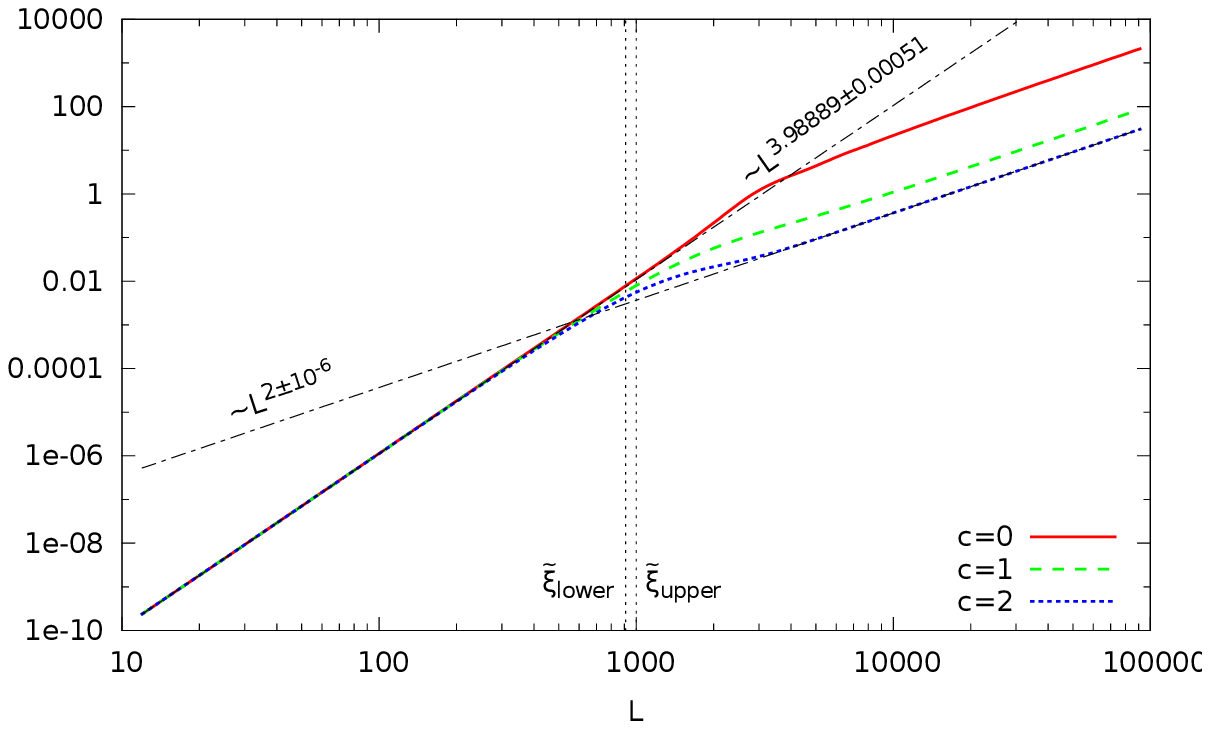}
\end{subfigure}
\begin{subfigure}[b]{0.48\textwidth}
\centering
\includegraphics[width=8cm,clip=on]{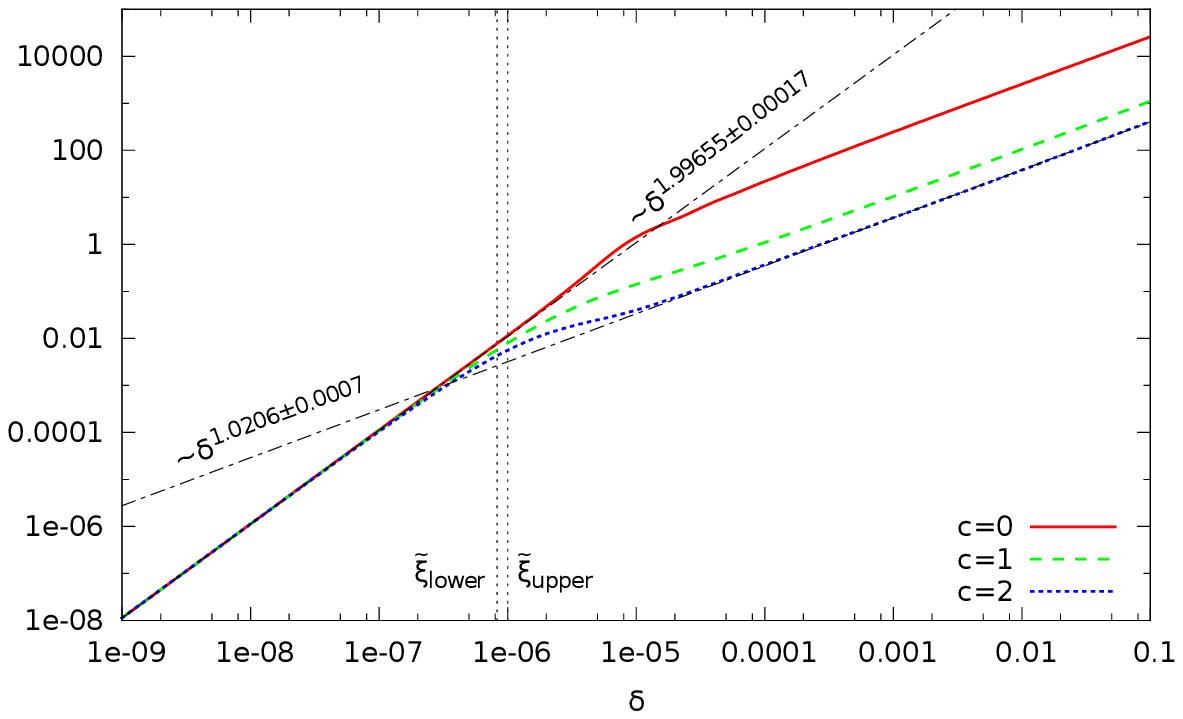}
\end{subfigure}
\caption{\label{2ddiff_lnF_v_L_ddmu1e-6_J1}\label{2ddiff_lnF_v_ddmu_L1000}(Color online) The antisymmetric model, critical point $\mu{=}2$ and $J{=}1$, $\delta {\equiv} \mu$.
Plots of $-\ln {\mathscr{F}}_{(1,0)}((2,1), \delta)$
for three values of $c$: $c{=}0$ -- red line, $c{=}1$ -- green line, $c{=}2$ -- blue line.
Left panel: plots of  $-\ln {\mathscr{F}}_{(1,0)}((2,1), 10^{-6})$ versus $L$.
Right panel: plots of $-\ln {\mathscr{F}}_{(1,0)}((2,1), \delta)$ versus $\delta$ for $L=10^3$.
Both plots are in doubly logarithmic scale and the black dashed-dotted straight lines indicate the power-law scaling. For more details see the text.}
\end{figure}

For the considered in this section critical points $\delta{\equiv}\mu$. In distinction to the symmetric model, the hierarchy of direction-dependent effective correlation lengths is not uniform in $J$ \cite{ajk-1}. However, for  $J{>}J_0$, with $J_0{\approx}1/4$, it is the same as in the symmetric case and is given in (\ref{tilde xi ineq}).
Consequently, for $J{>}J_0$ the lower and upper effective correlation lengths and the associated lower and upper deviations from the critical point are given by (\ref{tilde xi low upp}) and (\ref{delta low upp}), respectively. The critical exponents are direction independent,
$\nu_{\text{offdiag}}{=}\nu_{\text{diag}}{=}1/2$, and satisfy the condition $D\nu{<}2$.
In Fig.~\ref{2ddiff_lnF_v_L_ddmu1e-6_J1} we show plots of fidelity versus $L$ and $\delta$ in the particular case of $J{=}1$. Since in the scale of
Fig.~\ref{2ddiff_lnF_v_L_ddmu1e-6_J1}, ${\tilde{\xi}}_{\text{lower}}(\delta)$ and ${\tilde{\xi}}_{\text{upper}}(\delta)$ are quite close to each other, hence $\tilde{\delta}_{\text{lower}}(L)$ and $\tilde{\delta}_{\text{upper}}(L)$ are close as well, one can hardly distinguish two crossover regions, so the mezoscopic-system regime between them is very narrow. We can say that there is just one crossover region whose position is given by, say, ${\tilde{\xi}}_{\text{lower}}(\delta)$ (left panel) or by
$\tilde{\delta}_{\text{lower}}(L)$ (right panel). In the small-system regime of $L{<}{\tilde{\xi}}_{\text{lower}}(\delta)$ fidelity scales as $L^4$, which via (\ref{susc scaling 3}) gives $\nu=1/2$ in agreement with the exact value, while in the small-system regime of $\delta{<}\tilde{\delta}_{\text{lower}}(L)$ -- the standard $\delta^2$ scaling is observed. Then, in the macroscopic-system regime of $L{>}{\tilde{\xi}}_{\text{lower}}(\delta)$ the standard scaling $L^2$ holds, while in the macroscopic-system regime of $\delta{>}\tilde{\delta}_{\text{lower}}(L)$ fidelity scales as $\delta$, which via (\ref{fid scal}) reproduces again the exact value $\nu{=}1/2$. Thus, in the case of antisymmetric model the predictions of critical-scaling theory of fidelity match very well the exact results.

\section{\label{summ} Summary}

Hitherto, the quantum-critical-scaling theory of quantum fidelity has been verified in a few 1D models, which are equivalent to 1D lattice-fermion gases with gauge-symmetry breaking interaction: an Ising chain in a transverse magnetic field \cite{rams PRL 11}, an anisotropic XY chain in a transverse magnetic field \cite{rams PRA 11}, and the 1D version of the symmetric model considered in this paper \cite{ajk-2}. It has been demonstrated that this theory holds very well in neighborhoods of ordinary critical points, that is those critical points that are characterized by the unique correlation length and the associated universal critical index. In all those cases, the quantum-critical-scaling theory of quantum fidelity enables one to determine the values of correlation lengths and critical indices. Note however that in all the cited models the condition $D\nu{<}2$ is satisfied for any critical point. Only in neighborhoods of multicritical points, where the correlation lengths and their critical indices are not unique, breaking of the laws of quantum-critical-scaling theory has been observed \cite{rams PRA 11},\cite{ajk-2}.

Our aim in this paper has been to accomplish a similar task of verification but for a higher-dimensional model. The novelty of such models, as compared to 1D ones, is that the correlation lengths and the corresponding critical exponents $\nu$ may depend on spatial directions and the values of $\nu$ may be so large that the crucial for quantum-critical-scaling theory condition $D\nu{<}2$ is violated--the opposite strict inequality is satisfied. In case of the considered here 2D quasifree lattice-fermion models, we show that this is indeed the case. To this end we provide analytic formulae for the large-distance asymptotic behavior of two-point correlation functions, the correlation lengths, and the values of the corresponding exponents $\nu$ (a more comprehensive analysis can be found in \cite{ajk-1}). In all those cases where we failed to obtain analytic results, suitable numerical results are provided.

Some natural questions can be raised: how a multitude of correlation lengths and a multiplicity of critical indices at a given critical point, and or the violation of the condition $D\nu{<}2$ is reflected in the behavior of quantum fidelity?, is it possible, in those not encountered in 1D situations, to read off values of correlation lengths and critical indices from suitable plots of fidelity?

Below, we present in a compact form the results of our attempts to answer the raised questions.  All the data generated in our numerical calculations of fidelity $\ln {\mathscr{F}}_{\bs e}({\bs\lambda}_c, \delta)$ are contained in two kinds of plots,
$\ln {\mathscr{F}}_{\bs e}({\bs\lambda}_c, \delta)$ versus system's linear size $L$ and $\ln {\mathscr{F}}_{\bs e}({\bs\lambda}_c, \delta)$ versus a distance $\delta$ to a critical point, in a vicinity of that point. In the paper we present only representative excerpts of such plots.

The critical points of the considered models, and corresponding plots of fidelity, one can split into two groups: the ordinary critical points and multicritical ones.
Taking for granted the ubiquitous dependence of quantities of interest on spatial directions, let us call a critical point ordinary, if for a fixed spatial direction it is characterized by the unique correlation length. Then, in vicinities of ordinary critical points, the examples studied in previous sections show that if condition $D\nu{<}2$ holds true in any direction, then fidelity exhibits two crossover regions
(see Fig.~\ref{2d_lnF_v_L_mu0}, Fig.~\ref{2d_lnF_v_L_J0ep}, and Fig.~\ref{2ddiff_lnF_v_L_ddmu1e-6_J1}). That located at a smaller $L$ (or $\delta$) separates the small-system (quasi-critical) regime from a mezoscopic one. Its position is given by the lower effective correlation length ${\tilde{\xi}}_{\text{lower}}(\delta)$ (or the lower effective deviation ${\tilde{\delta}}_{\text{lower}}(L)$), which is the minimum over all the spatial directions of effective correlation lengths (or effective deviations). In the small-system regime, that is well below ${\tilde{\xi}}_{\text{lower}}(\delta)$ (or  ${\tilde{\delta}}_{\text{lower}}(L)$), small-system scaling law (\ref{susc scaling 3}) is satisfied. The second crossover region separates the mezoscopic regime from the macroscopic-system (off-critical) regime and is located by the upper effective correlation length ${\tilde{\xi}}_{\text{upper}}(\delta)$ (or the upper effective deviation ${\tilde{\delta}}_{\text{upper}}(L)$), which is the maximum over all the spatial directions of effective correlation lengths (or effective deviations). In the macroscopic-system regime, that is well above ${\tilde{\xi}}_{\text{upper}}(\delta)$ (or ${\tilde{\delta}}_{\text{upper}}(L)$), macroscopic-system scaling law (\ref{fid scal}) holds true.
If condition $D\nu{<}2$ is satisfied only for $\nu$ corresponding to ${\tilde{\xi}}_{\text{lower}}(\delta)$ (or to ${\tilde{\xi}}_{\text{upper}}(\delta)$), then only small-system scaling (\ref{susc scaling 3}) (macroscopic-system scaling (\ref{fid scal})) is obeyed by $\ln {\mathscr{F}}_{\bs e}({\bs\lambda}_c, \delta)$ for sufficiently large $L$  or $\delta$ but well below ${\tilde{\xi}}_{\text{lower}}(\delta)$
or ${\tilde{\delta}}_{\text{lower}}(L)$, respectively (for $L$ or $\delta$ well above ${\tilde{\xi}}_{\text{upper}}(\delta)$ or ${\tilde{\delta}}_{\text{upper}}(L)$).

After that, if in a vicinity of an ordinary critical point condition $D\nu{<}2$ is violated in any spatial direction (see Fig.~\ref{2d_lnF_v_L_J0}),
then $\ln {\mathscr{F}}_{\bs e}({\bs\lambda}_c, \delta)$ may scale, possibly in the generalized sense, according to a power law, and exhibit some crossovers, but neither the positions of crossovers are given by some ${\tilde{\xi}}_{n}(\delta)$ nor the scaling law,
(\ref{susc scaling 3})---for sufficiently small $L$ or $\delta$ and (\ref{fid scal})---for sufficiently large $L$ or $\delta$, is satisfied.
In this case, of note is also the superextensive behaviour of $-\ln {\mathscr{F}}_{\bs e}({\bs\lambda}_c, \delta)$,
%see section \ref{mu0},
which seems to defy the scaling arguments, according to which that should not be the case \cite{venuti zanardi 07, schwandt 09, albuquerque 10}. This curious phenomenon perhaps merits some further investigations into the properties of fidelity itself.

Finally, in the case of the multicritical point of the symmetric model we have faced a novel situation, never before encountered in such studies: the obtained critical indices $\nu$ satisfy the strict inequality $D\nu{>}2$. The behavior of $\ln {\mathscr{F}}_{\bs e}({\bs\lambda}_c, \delta)$ does allow to read off neither correlation lengths nor critical indices (see Fig.~\ref{2d_lnF_v_L_mcp}).

\begin{center}
{\bf Acknowledgements}\\
The presented studies have been supported by the University of Wroc\l aw through the projects Nr 1354/M/IFT/13 and Nr 1009/S/IFT/14.
\end{center}

\end{document}